\documentclass{article}
\usepackage[utf8]{inputenc}
\usepackage{macros}
\usepackage{pythonhighlight}

\usepackage{graphicx}
\graphicspath{ {./images/} }

\newcommand{\OPT}{\mathsf{OPT}}

\DeclareMathOperator{\diam}{diam}

\title{Coresets for Multiple $\ell_p$ Regression}
\author{
David P. Woodruff \\ Carnegie Mellon University \\ \texttt{dwoodruf@cs.cmu.edu} \and 
Taisuke Yasuda \\ Carnegie Mellon University \\ \texttt{taisukey@cs.cmu.edu}
}

\begin{document}

\maketitle

\thispagestyle{empty}
\begin{abstract}
A \emph{coreset} of a dataset with $n$ examples and $d$ features is a weighted subset of examples that is sufficient for solving downstream data analytic tasks. Nearly optimal constructions of coresets for least squares and $\ell_p$ linear regression with a single response are known in prior work. However, for multiple $\ell_p$ regression where there can be $m$ responses, there are no known constructions with size sublinear in $m$. In this work, we construct coresets of size $\tilde O(\varepsilon^{-2}d)$ for $p<2$ and $\tilde O(\varepsilon^{-p}d^{p/2})$ for $p>2$ independently of $m$ (i.e., dimension-free) that approximate the multiple $\ell_p$ regression objective at every point in the domain up to $(1\pm\varepsilon)$ relative error. If we only need to preserve the minimizer subject to a subspace constraint, we improve these bounds by an $\varepsilon$ factor for all $p>1$. All of our bounds are nearly tight.

We give two application of our results. First, we settle the number of uniform samples needed to approximate $\ell_p$ Euclidean power means up to a $(1+\varepsilon)$ factor, showing that $\tilde\Theta(\varepsilon^{-2})$ samples for $p = 1$, $\tilde\Theta(\varepsilon^{-1})$ samples for $1 < p < 2$, and $\tilde\Theta(\varepsilon^{1-p})$ samples for $p>2$ is tight, answering a question of Cohen-Addad, Saulpic, and Schwiegelshohn. Second, we show that for $1<p<2$, every matrix has a subset of $\tilde O(\varepsilon^{-1}k)$ rows which spans a $(1+\varepsilon)$-approximately optimal $k$-dimensional subspace for $\ell_p$ subspace approximation, which is also nearly optimal.
\end{abstract}

\clearpage
\setcounter{page}{1}

\section{Introduction}

Least squares linear regression and $\ell_p$ linear regression are some of the most fundamental and practically valuable computational problems in statistics and optimization. In this problem, our input is an $n\times d$ matrix $\bfA\in\mathbb R^{n\times d}$ and a response vector $\bfb\in\mathbb R^n$, and our goal is to output an approximate minimizer $\hat\bfx\in\mathbb R^d$ such that
\begin{equation}
\label{eq:lp-regression}
    \norm{\bfA\hat\bfx-\bfb}_p^p \leq (1+\eps)\min_{\bfx\in\mathbb R^d}\norm{\bfA\bfx-\bfb}_p^p.
\end{equation}
Among the vast literature on $\ell_p$ regression, sampling algorithms and coresets, which are algorithms that select a weighted subset of the rows of $\bfA$ and $\bfb$ that suffice to solve $\ell_p$ regression, have played major roles in the development of efficient algorithms. That is, we seek a diagonal matrix $\bfS\in\mathbb R^{n\times n}$ with few non-zero entries, i.e., $\nnz(\bfS) \ll n$, such that the weighted subset of rows $\bfS\bfA$ and $\bfS\bfb$ are sufficient to compute a solution $\hat\bfx$ satisfying \eqref{eq:lp-regression}. We will often refer to $\nnz(\bfS)$ as the \emph{sample complexity}. We focus on approaches that construct $\bfS$ by i.i.d.\ sampling of each of the $n$ rows:

\begin{Definition}[$\ell_p$ sampling matrix]
\label{def:sampling-matrix}
Let $p\geq 1$. A random diagonal matrix $\bfS\in\mathbb R^{n\times n}$ is a \emph{random $\ell_p$ sampling matrix with sampling probabilities $\{q_i\}_{i=1}^n$} if for each $i\in[n]$, the $i$th diagonal entry is independently set to be
\[
    \bfS_{i,i} = \begin{cases}
        1 / q_i^{1/p} & \text{with probability $q_i$} \\
        0 & \text{otherwise}
    \end{cases}
\]
\end{Definition}

Two well-studied guarantees for $\bfS$ are \emph{strong coresets} and \emph{weak coresets}. Strong coresets refer to coresets that preserve the value of the objective function at \emph{every} point in the domain, while weak coresets only guarantee that the unconstrained minimizer is preserved. If we only care about solving the unconstrained $\ell_p$ regression problem, then weak coresets are sufficient to solve this problem, and it is known that weak coresets can be substantially smaller than strong coresets in certain settings \cite{MMWY2022}. On the other hand, strong coresets are necessary when the objective function must be evaluated at points away from the optimum, for example for constrained optimization problems.

\begin{Definition}[Strong coreset]
\label{def:strong-coreset}
We say that $\bfS$ is a \emph{strong coreset} if $\norm{\bfS(\bfA\bfx-\bfb)}_p^p = (1\pm\eps)\norm{\bfA\bfx-\bfb}_p^p$ simultaneously for every $\bfx\in\mathbb R^d$.
\end{Definition}

\begin{Definition}[Weak coreset]
\label{def:weak-coreset}
We say that $\bfS$ is a \emph{weak coreset} if
\[
    \norm{\bfA\hat\bfx-\bfb}_p^p \leq (1+\eps)\min_{\bfx\in\mathbb R^d}\norm{\bfA\bfx-\bfb}_p^p
\]
for $\hat\bfx = \arg\min_{\bfx\in\mathbb R^d}\norm{\bfS(\bfA\bfx-\bfb)}_p^p$.
\end{Definition}

The efficient construction of coresets for $\ell_p$ regression has been studied in a long line of work \cite{Cla2005, DMM2006, DMM2006b, DDHKM2009} culminating in the $\ell_p$ Lewis weight sampling algorithm \cite{Lew1978, BLM1989, Tal1990, LT1991, Tal1995, SZ2001, CP2015, WY2023a}, which gives an algorithm that constructs a strong coreset $\bfS$ with
\[
    \nnz(\bfS) = \begin{cases}
        \tilde O(\eps^{-2}d) & p \leq 2 \\
        \tilde O(\eps^{-2}d^{p/2}) & p > 2
    \end{cases}.
\]
A related line of work in the \emph{active $\ell_p$ regression} setting shows that weak coresets for $\ell_p$ regression with
\[
    \nnz(\bfS) = \begin{cases}
        \tilde O(\eps^{-2}d) & p = 1 \\
        \tilde O(\eps^{-1}d) & 1 < p < 2 \\
        O(\eps^{-1}d) & p = 2 \\
        \tilde O(\eps^{-(p-1)}d^{p/2}) & p > 2
    \end{cases}
\]
can be constructed even without knowing $\bfb$ \cite{CP2019, CD2021, PPP2021, MMWY2022, WY2023b}. Note that these bounds strictly improve over the strong coreset guarantees of $\ell_p$ Lewis weight sampling for $1 < p < 3$.

\subsection{Multiple \texorpdfstring{$\ell_p$}{lp} regression}

It is often the case that we are interested in more than just one target $\bfb$ to predict, and in general, we may wish to simultaneously fit $m$ target vectors that are given by a matrix $\bfB\in\mathbb R^{n\times m}$ and solve the minimization problem
\[
    \min_{\bfX\in\mathbb R^{d\times m}}\norm*{\bfA\bfX - \bfB}_{p,p}^p = \min_{\bfX\in\mathbb R^{d\times m}}\sum_{j=1}^m \norm*{\bfA\bfX\bfe_j - \bfB\bfe_j}_p^p
\]
This is known as the \emph{multiple response $\ell_p$ regression} problem, or simply the \emph{multiple $\ell_p$ regression} problem, and is the focus of the present work.

\subsubsection{Coreset constructions for \texorpdfstring{$p=2$}{p=2}}

For $p = 2$, the construction of strong coresets for the multiple response problem follows almost immediately from strong coresets for the single response problem due to orthogonality and the Pythagorean theorem, and we can construct $\bfS$ such that
\[
    \norm{\bfS(\bfA\bfX-\bfB)}_F^2 = (1\pm\eps)\norm{\bfA\bfX-\bfB}_F^2
\]
with $\nnz(\bfS) = \tilde O(\eps^{-2}d)$ samples. Indeed, assume without loss of generality that $\bfA$ has orthogonal columns, and suppose that $\bfS$ satisfies 
\begin{itemize}
    \item $\norm{\bfS\bfA\bfx}_2^2 = (1\pm\eps)\norm{\bfA\bfx}_2^2$ for every $\bfx\in\mathbb R^d$ (i.e., $\bfS$ is a subspace embedding)
    \item $\norm{\bfS(\bfA\bfX^*-\bfB)}_F^2 = (1\pm\eps)\norm{\bfA\bfX^*-\bfB}_F^2$ where $\bfX^*$ is the optimal minimizer
    \item $\norm{\bfA^\top\bfS^\top\bfS(\bfA\bfX^*-\bfB)}_F^2 \leq (\eps^2/d)\norm{\bfA}_F^2\norm{\bfA\bfX^*-\bfB}_F^2 = \eps^2\norm{\bfA\bfX^*-\bfB}_F^2$
\end{itemize}
Then, the following argument of Section 7.5 of \cite{CW2013} shows that $\bfS$ is a strong coreset. Indeed,
\begin{align*}
    \norm{\bfS(\bfA\bfX-\bfB)}_F^2 &= \norm{\bfS\bfA(\bfX-\bfX^*)}_F^2 + \norm{\bfS(\bfA\bfX^*-\bfB)}_F^2 + 2\tr\parens*{(\bfX-\bfX^*)^\top\bfA^\top\bfS^\top\bfS(\bfA\bfX^*-\bfB)}
\end{align*}
by expanding the square, and the inner product term is bounded by
\begin{align*}
    \abs*{\tr\parens*{(\bfX-\bfX^*)^\top\bfA^\top\bfS^\top\bfS(\bfA\bfX^*-\bfB)}} &\leq \norm{\bfX-\bfX^*}_F\norm{\bfA^\top\bfS^\top\bfS(\bfA\bfX^*-\bfB)}_F \\
    &\leq \eps\norm{\bfA(\bfX-\bfX^*)}_F\norm{\bfA\bfX^*-\bfB}_F \\
    &\leq\eps\norm{\bfA\bfX-\bfB}_F^2
\end{align*}
and $\bfS$ also preserves the quantities $ \norm{\bfS\bfA(\bfX-\bfX^*)}_F^2$ and $\norm{\bfS(\bfA\bfX^*-\bfB)}_F^2$ up to $(1\pm\eps)$ relative error. A similar trick is available in the weak coreset setting (see, e.g., Section 3.1 of \cite{CNW2016}), which gives a bound of $\nnz(\bfS) = \tilde O(\eps^{-1}d)$ for this guarantee. Unfortunately, almost every step in the above argument uses special properties of the $\ell_2$ norm that are not available for the $\ell_p$ norm, and thus we will need completely different arguments to handle $p\neq 2$.

\subsubsection{Challenges for \texorpdfstring{$p\neq 2$}{p != 2}}

If we desire only weak coresets, then prior results on active $\ell_p$ regression in fact almost immediately provide a solution. These results show that a weak coreset $\bfS$ for the single response $\ell_p$ regression problem can be constructed independently of $\bfb$, and with the dependence of $\nnz(\bfS)$ on the failure probability $\delta$ being polylogarithmic. Thus by setting the failure rate to $\delta = 1/10m$, we can simultaneously solve every column of $\bfB$ independently with overall probability at least $9/10$.

For strong coresets, however, such a column-wise strategy must be implemented carefully. If we consider constructing a strong coreset for a single column $j\in[m]$, then the sampling probabilities now depend on the target vector $\bfB\bfe_j$, so the sampling complexity would need to scale as $m$ rather than $\poly\log(m)$ as in the previous upper bound weak coresets. On the other hand, another natural strategy is to mimic the strategy for the $p = 2$ case and take the sampling probabilities to only guarantee an $\ell_p$ subspace embedding for the column space of $\bfA$ and that $q_i \geq \norm{\bfe_i^\top\bfB^*}_p^p / \norm{\bfB^*}_{p,p}^p$ for $\bfB^* \coloneqq \bfA\bfX^*-\bfB$. This is a reasonable choice of sampling probabilities, and indeed it is not hard to see that
\[
    \norm{\bfS(\bfA\bfX-\bfB)}_{p,p}^p = (1\pm\eps)\norm{\bfA\bfX-\bfB}_{p,p}^p
\]
for any fixed $\bfX\in\mathbb R^{d\times m}$ with only $\nnz(\bfS) = \tilde O(\eps^{-2}d)$ samples for $p<2$ and $\nnz(\bfS) = \tilde O(\eps^{-2}d^{p/2})$ samples for $p>2$ via a Bernstein tail bound. However, it is unclear how to extend a guarantee for any single $\bfX\in\mathbb R^{d\times m}$ to a guarantee simultaneously for \emph{all $\bfX\in\mathbb R^{d\times m}$}. Although the dependence on the failure rate $\delta$ is logarithmic, a net argument, or even more sophisticated chaining arguments, over the possible choices of $\bfX\in\mathbb R^{d\times m}$ seem to require a union bound over sets of size $\exp(dm)$, thus again introducing a linear dependence on $m$ in the sample complexity $\nnz(\bfS)$. As we show, a careful blend of these two ideas will be necessary to obtain our strong coreset result.

\subsection{Strong coresets for multiple \texorpdfstring{$\ell_p$}{lp} regression}

Our first main result is the first construction of strong coresets for multiple $\ell_p$ regression that is independent of $m$.

\begin{Theorem}[Strong coresets for multiple $\ell_p$ regression]
\label{thm:strong-coreset-informal}
Let $\bfA\in\mathbb R^{n\times d}$, $\bfB\in\mathbb R^{n\times m}$, and $p\geq 1$. There is an algorithm which constructs $\bfS$ with
\[
    \nnz(\bfS) = \begin{dcases}
        \frac{O(d)}{\eps^2}\bracks*{(\log d)^2\log\frac{d}{\eps} + \log\frac{1}{\delta}} & 1\leq p < 2 \\
        \frac{O(d^{p/2})}{\eps^p}\bracks*{(\log d)^2\log\frac{d}{\eps} + \log\frac{1}{\delta}} & p > 2 \\
    \end{dcases}
\]
such that with probability at least $1-\delta$,
\[
    \norm{\bfS(\bfA\bfX-\bfB)}_{p,p}^p = (1\pm\eps)\norm{\bfA\bfX-\bfB}_{p,p}^p
\]
simultaneously for every $\bfX\in\mathbb R^{d\times m}$. Furthermore, $\bfS$ can be constructed in $\tilde O(\nnz(\bfA) + \nnz(\bfB) + \poly(d))$ time.
\end{Theorem}

We achieve a nearly optimal dependence on $d$ and $\eps$, as we show that $\Omega(d^{p/2}/\eps^p)$ rows are necessary for strong coresets in Theorem \ref{thm:strong-coreset-lower-bound} for $p>2$, while it is known that $\tilde\Omega(d/\eps^2)$ rows are necessary even for $m = 1$ for $p < 2$ \cite{LWW2021}. We note that our upper bound shows that multiple $\ell_p$ regression is as easy as single response $\ell_p$ regression for $p<2$, while our lower bound demonstrates an interesting separation between the two for $p>2$.

\subsubsection{Initial \texorpdfstring{$\log m$}{log m} bound}

Our main technique is to generalize the ``partition by sensitivity'' technique introduced in the active $\ell_p$ regression work of \cite{MMWY2022} and show how this can be applied to the strong coreset setting. We describe the idea for the case of $p<2$, as the case of $p>2$ is analogous.

In the active $\ell_p$ regression setting, we must show that we can design sampling algorithms that preserve the objective function value, even if we do not know the target vector $\bfb$. In this setting, one of the main observations of \cite{MMWY2022} is that even though we cannot preserve $\norm{\bfA\bfx-\bfb}_p^p$ itself, we can actually preserve the difference $\norm{\bfA\bfx-\bfb}_p^p - \norm{\bfb}_p^p$, if $\norm{\bfb}_p^p = O(\OPT^p)$ which is without loss of generality. To see this idea, assume (without loss of generality due to \cite{DDHKM2009}) that we restrict our attention to $\norm{\bfA\bfx}_p^p = O(\OPT^p)$. Then, the analysis of \cite{MMWY2022} proceeds by partitioning the coordinates of $\bfb$ into two sets, those such that $\abs{\bfb(i)}^p$ is larger than $\eps^{-p}\bfw_i \OPT^p$ and those that are smaller than this threshold, where $\bfw_i$ is the $i$-th $\ell_p$ Lewis weight of $\bfA$. It is known that $\bfw_i$ bounds the sensitivities of $\bfA$, that is, $\abs{[\bfA\bfx](i)}^p \leq \bfw_i \norm{\bfA\bfx}_p^p$ so it follows that for any $\abs{\bfb(i)}^p \geq \eps^{-p}\bfw_i \OPT^p$, we have that
\[
    \abs*{\abs{[\bfA\bfx-\bfb](i)}^p - \abs{\bfb(i)}^p} = O(\eps)\abs{\bfb(i)}^p
\]
for any $\bfx\in\mathbb R^d$ with $\norm{\bfA\bfx}_p^p = O(\OPT^p)$. On the other hand, if $\abs{\bfb(i)}^p \leq \eps^{-p}\bfw_i \OPT^p$, then we have by the triangle inequality that
\[
    \abs*{\abs{[\bfA\bfx-\bfb](i)}^p - \abs{\bfb(i)}^p} \leq O(\eps^{-p})\bfw_i \OPT^p.
\]
Thus, up to an additive $O(\eps)(\norm{\bfS\bfb}_p^p + \norm{\bfb}_p^p)$ error, $\abs*{\abs{[\bfA\bfx-\bfb](i)}^p - \abs{\bfb(i)}^p}$ has sensitivities which are controlled by the $\ell_p$ Lewis weights of $\bfA$. This allows one to show that sampling by the $\ell_p$ Lewis weights of $\bfA$ preserves $\norm{\bfA\bfx-\bfb}_p^p - \norm{\bfb}_p^p$ for all $\norm{\bfA\bfx}_p^p = O(\OPT^p)$.

In order to apply this idea to the strong coreset setting, we generalize the above argument to multiple scales. That is, we replace $\OPT^p$ by an arbitrary scale $R\geq \norm{\bfb}_p^p$, and show that for every $\norm{\bfA\bfx}_p^p \leq O(R)$ that
\begin{align*}
    \abs*{\parens*{\norm{\bfS(\bfA\bfx-\bfb)}_p^p - \norm{\bfS\bfb}_p^p} - \parens*{\norm{\bfA\bfx-\bfb}_p^p - \norm{\bfb}_p^p}} \leq \eps (R + \norm{\bfS\bfb}_p^p)
\end{align*}
Finally, we can generalize this to the following guarantee by union bounding over finitely many scales $R$, which holds for every $\bfx\in\mathbb R^d$:
\begin{equation}
\label{eq:lewis-diff}
\begin{aligned}
    \abs*{\parens*{\norm{\bfS(\bfA\bfx-\bfb)}_p^p - \norm{\bfS\bfb}_p^p} - \parens*{\norm{\bfA\bfx-\bfb}_p^p - \norm{\bfb}_p^p}} \leq \eps \parens*{\norm{\bfb}_p^p + \norm{\bfS\bfb}_p^p + \norm{\bfA\bfx}_p^p}
\end{aligned}
\end{equation}
This guarantee is in a form that can be summed over the $m$ columns of $\bfB$. Thus, if a $\log m$ dependence is admissible, then we can apply the above result with failure probability $1/10m$, union bound over the $m$ columns, and sum the results to obtain
\begin{align*}
    \left\lvert (\norm{\bfS(\bfA\bfX-\bfB)}_{p,p}^p-\norm{\bfS\bfB}_{p,p}^p)  - (\norm{\bfA\bfX-\bfB}_{p,p}^p-\norm{\bfB}_{p,p}^p) \right\rvert \leq \eps\parens*{\norm{\bfB}_{p,p}^p + \norm{\bfS\bfB}_{p,p}^p + \norm{\bfA\bfX}_{p,p}^p}.
\end{align*}
Now suppose that we additionally have
\begin{itemize}
    \item $\norm{\bfS\bfB}_{p,p}^p = (1\pm\eps)\norm{\bfB}_{p,p}^p$
    \item $\norm{\bfB}_{p,p}^p = O(\OPT^p)$ (which is without loss of generality by subtracting an $O(1)$-optimal solution)
\end{itemize}
Then, we have
\begin{align*}
    \norm{\bfS(\bfA\bfX-\bfB)}_{p,p}^p &= \norm{\bfA\bfX-\bfB}_{p,p}^p-\norm{\bfB}_{p,p}^p+\norm{\bfS\bfB}_{p,p}^p \pm O(\eps)\parens*{\norm{\bfB}_{p,p}^p + \norm{\bfA\bfX}_{p,p}^p} \\
    &= \norm{\bfA\bfX-\bfB}_{p,p}^p\pm\eps\norm{\bfB}_{p,p}^p \pm O(\eps)\parens*{\norm{\bfB}_{p,p}^p + \norm{\bfA\bfX}_{p,p}^p} \\
    &= \norm{\bfA\bfX-\bfB}_{p,p}^p \pm O(\eps) \norm{\bfA\bfX-\bfB}_{p,p}^p
\end{align*}
so we indeed have a strong coreset as desired.

\subsubsection{Removing the \texorpdfstring{$m$}{m} dependence}

Next, we show how to completely remove the $m$ dependence, which requires additional ideas. When applying \eqref{eq:lewis-diff} to each of the $m$ columns, suppose that we set the failure probability to $\poly(\eps\delta)$ instead of $O(1/m)$. Then, this guarantee will hold for a $1-\poly(\eps\delta)$ fraction of ``good'' columns, for which we can obtain $(1\pm\eps)$ approximations. On the remaining $\poly(\eps\delta)$ fraction of ``bad'' columns, note that the mass of $\bfB$ on these columns is at most $\poly(\eps\delta)\norm{\bfB}_{p,p}^p$ with probability $1-\delta$ by Markov's inequality. Then on these columns, $\norm{\bfS(\bfA\bfX-\bfB)\bfe_j}_p$ is just $\norm{\bfS\bfA\bfX\bfe_j}_p$ up to a small total additive error of $\poly(\eps\delta)\norm{\bfB}_{p,p}^p$. In turn, we have that $\norm{\bfS\bfA\bfX\bfe_j}_p = (1\pm\eps)\norm{\bfA\bfX\bfe_j}_p$ by using that $\bfS$ is an $\ell_p$ subspace embedding. Thus, by combining with the $(1\pm\eps)$ approximation on the rest of the ``good'' columns, we can still ensure that $\norm{\bfS(\bfA\bfX-\bfB)}_{p,p} = (1\pm\eps)\norm{\bfA\bfX-\bfB}_{p,p}$.

\subsection{Weak coresets for multiple \texorpdfstring{$\ell_p$}{lp} regression}

In the weak coreset setting, we consider a generalized multiple $\ell_p$ regression problem, where we are given a design matrix $\bfA\in\mathbb R^{n\times d}$, an ``embedding'' $\bfG\in\mathbb R^{t\times m}$, and a target matrix $\bfB\in\mathbb R^{n\times m}$, and we wish to approximately minimize the objective function $\norm{\bfA\bfX\bfG-\bfB}_{p,p}$. 

As noted previously, for multiple $\ell_p$ regression without an embedding (i.e., $\bfG = \bfI_t$) the construction of weak coresets follows relatively straightforwardly by applying active $\ell_p$ regression results along each column. However, this strategy fails when we must additionally handle the embedding matrix $\bfG$, as this constraint couples the columns of $\bfA\bfX$ together. Furthermore, we argue that handling the embedding $\bfG$ is substantially more interesting that the unconstrained case. Indeed, as we see later in Sections \ref{sec:power-means-intro} and \ref{sec:subspace-approx-intro}, the incorporation of the embedding $\bfG$ will allow us to handle interesting extensions of our results to settings beyond the entrywise $\ell_p$ norm via the use of a linear embedding into this norm. We will denote the optimal value as
\[
    \OPT \coloneqq \min_{\bfX\in\mathbb R^{d\times t}} \norm{\bfA\bfX\bfG-\bfB}_{p,p}
\]
and let $\bfX^*$ denote the matrix achieving this optimum unless otherwise noted. We will prove the following result:

\begin{Theorem}[Weak coresets for multiple $\ell_p$ regression]
\label{thm:weak-coreset-informal}
Let $\bfA\in\mathbb R^{n\times d}$, $\bfG\in\mathbb R^{t\times m}$, $\bfB\in\mathbb R^{n\times m}$, and $1 \leq p < \infty$. There is an algorithm which constructs $\bfS$ independently of $\bfB$ with
\[
    \nnz(\bfS) = \frac{O(d)}{\eps^2 \delta^2}\bracks*{(\log d)^2\log \frac{d}{\eps} + \log\frac{1}{\delta}}\parens*{\log\log\frac1\eps}^2
\]
for $p = 1$,
\[
    \nnz(\bfS) = \frac{O(d)}{\eps \delta^2}\bracks*{(\log d)^2\log \frac{d}{\eps} + \log\frac{1}{\delta}}\parens*{\log\log\frac1\eps}^2
\]
for $1 < p < 2$, and
\[
    \nnz(\bfS) = \frac{O(d^{p/2})}{\eps^{p-1} \delta^p}\bracks*{(\log d)^2\log \frac{d}{\eps} + \log\frac{1}{\delta}} \parens*{\log\log\frac1\eps}^p
\]
for $p > 2$ such that with probability at least $1-\delta$, for any $\hat\bfX\in\mathbb R^{d\times t}$ such that
\[
    \norm{\bfS(\bfA\hat\bfX\bfG-\bfB)}_{p,p}^p \leq (1+\eps)\min_{\bfX\in\mathbb R^{d\times t}} \norm{\bfS(\bfA\bfX\bfG-\bfB)}_{p,p}^p,
\]
we have
\[
    \norm{\bfA\hat\bfX\bfG-\bfB}_{p,p}^p \leq (1+O(\eps))\min_{\bfX\in\mathbb R^{d\times t}} \norm{\bfA\bfX\bfG-\bfB}_{p,p}^p.
\]
Conditioned on the event that $\norm{\bfS(\bfA\bfX^*\bfG-\bfB)}_{p,p}^p = O(\norm{\bfA\bfX^*\bfG-\bfB}_{p,p}^p)$ for the global optimizer $\bfX^*$, the dependence on $\delta$ can be replaced by a single $\log\frac1\delta$ factor and the $\poly(\log\log\frac1\eps)$ factor can be removed. Furthermore, $\bfS$ can be constructed in $\tilde O(\nnz(\bfA) + d^\omega)$ time.
\end{Theorem}

We achieve a nearly optimal dependence on $d$ and $\eps$, as we show that $\Omega(d^{p/2}/\eps^{p-1})$ rows are necessary for weak coresets in Theorem \ref{thm:weak-coreset-lower-bound} for $p>2$. Our weak coreset upper bound result together with our strong coreset lower bound of Theorem \ref{thm:strong-coreset-lower-bound} shows a tight $\eps$ factor separation between the two coreset guarantees.

Note that in the statement of Theorem \ref{thm:weak-coreset-informal}, the dependence on the failure rate $\delta$ is polynomial. This is in fact necessary if we restrict our algorithm to be of the form of ``sample-and-solve'' algorithms whose sampling matrices $\bfS$ do not depend on $\bfB$, as demonstrated in a lower bound result of Theorem 12.8 of \cite{MMWY2022}. The only reason why this dependence becomes necessary in the analysis of the upper bound is that $\norm{\bfS(\bfA\bfX^*\bfG-\bfB)}_{p,p}^p$ may be as large as $O\parens*{\frac1\delta}\norm{\bfA\bfX^*\bfG-\bfB}_{p,p}^p$ with probability at least $\delta$, and this is the source of the hardness result of Theorem 12.8 of  \cite{MMWY2022} as well. This is a mild problem and can be easily circumvented in one of two ways. The first is to simply allow the algorithm to incorporate the row norms of $\bfB$ into the sampling probabilities just as in Theorem \ref{thm:strong-coreset-informal}. However, this would not give an active regression algorithm that makes only polylogarithmic in $\delta$ many queries. If we wish for such an active regression algorithm, then we can follow \cite{MMWY2022} and consider the following two-stage procedure. First, we can obtain a constant factor solution $\hat\bfX$ with a polylogarithmic dependence on $\delta$ by employing a ``median''-like procedure (see Section 3.1 of \cite{MMWY2022}). Then, we can run $\log\frac1\delta$ copies of the algorithm, each of which succeeds with probability $1-\delta$. Then, we can sort the runs by their estimates $\norm{\bfS(\bfA\hat\bfX\bfG-\bfB)}_{p,p}^p$ and discard half of the runs with the highest values of $\norm{\bfS(\bfA\hat\bfX\bfG-\bfB)}_{p,p}^p$. This guarantees that the remaining runs have $\norm{\bfS(\bfA\bfX^*\bfG-\bfB)}_{p,p}^p = O(1)\norm{\bfA\bfX^*\bfG-\bfB}_{p,p}^p$ with probability at least $1-\delta$, which is enough for the rest of the argument to go through with only a polylogarithmic dependence on $\delta$.

\subsection{Applications: sublinear algorithms for Euclidean power means}
\label{sec:power-means-intro}

Our first application of our results on coresets for multiple $\ell_p$ regression is on designing coresets for the Euclidean power means problem. In this problem, we are given as input a set of $n$ points $\{\bfb_i\}_{i=1}^n\subseteq\mathbb R^t$, and we wish to find a center $\hat\bfx\in\mathbb R^t$ that minimizes the sum of the Euclidean distances to $\hat\bfx$, raised to the power $p$. That is, we seek to minimize the objective function given by
\[
    \sum_{i=1}^n \norm{\bfx - \bfb_i}_2^p = \norm{\mathbf{1}\bfx^\top-\bfB}_{p,2}^p
\]
where $\mathbf{1}$ is the $n\times 1$ matrix of all ones, $\bfB\in\mathbb R^{n\times t}$ is the matrix with $\bfb_i$ in its $n$ rows, and $\norm{\cdot}_{p,2}$ is the $(p,2)$-norm of a matrix given by the $\ell_p$ norm of the Euclidean norm of the rows. This is a fundamental problem which generalizes the well-studied problems of the mean ($p = 2$), geometric median ($p = 1$), and minimum enclosing balls ($p = \infty$). Coresets and sampling algorithms for this problem were recently studied by \cite{CSS2021b}, who showed that a uniform sample of $\tilde O(\eps^{-(p+3)})$ points suffices to output a center $\hat\bfx\in\mathbb R^t$ such that
\[
    \norm{\mathbf{1}\hat\bfx^\top-\bfB}_{p,2}^p \leq (1+\eps)\min_{\bfx\in\mathbb R^t}\norm{\mathbf{1}\bfx^\top-\bfB}_{p,2}^p = (1+\eps)\OPT^p.
\]
In comparison to the upper bounds, the lower bounds given by \cite{CSS2021b} was $\Omega(\eps^{-(p-1)})$ which is off by an $\eps^4$ factor compared to the upper bound, which was improved to $\Omega(\eps^{-1})$ for $1<p<2$ by \cite{MMWY2022} and $\Omega(\eps^{-2})$ for $p=1$ by \cite{CD2021, PPP2021}.

One of the main open questions highlighted by the work of \cite{CSS2021b} is to obtain tight bounds for this problem: how many uniform samples are necessary and sufficient to output a $(1+\eps)$-approximate solution to the Euclidean power means problem. Our main contribution is a nearly optimal algorithm which matches the lower bounds of \cite{CD2021, PPP2021, CSS2021b, MMWY2022}.

\begin{restatable}{Theorem}{PowerMeans}
\label{thm:power-means}
Let $\{\bfb_i\}_{i=1}^n\subseteq\mathbb R^d$. Then, there is a sublinear algorithm which uniformly samples at most
\[
    s = \begin{cases}
        O(\eps^{-2})\parens*{\log\frac1\eps + \log\frac1\delta}\log\frac1\delta & p = 1 \\
        O(\eps^{-1})\parens*{\log\frac1\eps + \log\frac1\delta}\log\frac1\delta & 1 < p \leq 2 \\
        O(\eps^{1-p})\parens*{\log\frac1\eps + \log\frac1\delta}\log\frac1\delta & 2 < p < \infty
    \end{cases}
\]
rows $\bfb_i$ and outputs a center $\hat\bfx$ such that
\[
    \sum_{i=1}^n \norm{\hat\bfx-\bfb_i}_2^p \leq (1+\eps)\min_{\bfx\in\mathbb R^d} \sum_{i=1}^n \norm{\bfx-\bfb_i}_2^p
\]
with probability at least $1-\delta$.
\end{restatable}

To apply the techniques developed in this work to the Euclidean power means problem, we need to embed the $(p,2)$-norm into the entrywise $\ell_p$ norm. To make this reduction, we use a classic result of Dvoretzky and Milman \cite{Dvo1961, Mil1971}, which shows that a random subspace of a normed space is approximately Euclidean. We will need the following version of this result for $\ell_p$ norms:

\begin{Theorem}[Dvoretzky's theorem for $\ell_p$ norms \cite{FLM1977, PVZ2017}]
\label{thm:dvoretzky}
Let $p\geq 1$ and $0 < \eps < 1/p$. Let $n\geq O(\max\{\eps^{-2}k, \eps^{-1}k^{p/2})$, and let $\bfG\in\mathbb R^{n\times k}$ be an i.i.d.\ random Gaussian matrix. Then, with probability at least $2/3$, $\norm*{\bfG\bfx}_p^p = (1\pm\eps) n\norm*{\bfx}_2^p$ for every $\bfx\in\mathbb R^k$.
\end{Theorem}

Note then that if $\bfG$ is an appropriately scaled random Gaussian matrix, then we have that
\[
    \norm{\mathbf{1}\bfx^\top-\bfB}_{p,2}^p = (1\pm\eps)\norm{\mathbf{1}\bfx^\top\bfG-\bfB\bfG}_{p,p}^p
\]
by the above result. We may now note that the latter optimization problem is exactly of the form of an embedded $\ell_p$ regression problem, and thus our weak coreset results immediately apply to this problem. In fact, handling this Dvoretzky embedding is our main motivation for studying the $\ell_p$ regression problem with the embedding. We also note that similar reductions are possible by making use of other linear embeddings between $\ell_p$ norms \cite{WW2019, LWY2021, LLW2023}. The full argument is given in Appendix \ref{sec:power-means}.

In addition to sharpening the bound of \cite{CSS2021b} to optimality, we note that our techniques, both algorithmically and in the analysis, are simpler than the prior work of \cite{CSS2021b}. The previous algorithm required partitioning the dataset into ``rings'' of points with similar costs and preprocessing these rings. Furthermore, the analysis uses a specially designed chaining argument with custom net constructions that require terminal Johnson--Lindenstrauss embeddings. On the other hand, our algorithm simply runs multiple instances of a ``sample-and-solve'' algorithm, where the run with lowest sampled mass is kept. Furthermore, the analysis largely builds on existing net constructions for $\ell_p$ regression, and does not need terminal embeddings. In fact, our proof for the power means problem only need $\ell_p$ regression net constructions in $d = 1$ dimensions due to our use of Dvoretzky's theorem, which avoids the sophisticated constructions of \cite{BLM1989} for large $d$ and only needs a standard volume argument (Remark \ref{rem:simple-net}). Our partition of sensitivity can also be thought of as a coarse notion of rings, where we only consider two classes of costs, ``big'' and ``small'', whereas prior work requires finer a classification of points into rings of points whose costs are related up to a constant factor.

\subsection{Applications: spanning coresets for \texorpdfstring{$\ell_p$}{lp} subspace approximation}
\label{sec:subspace-approx-intro}

As a second application of our results, we give the first construction of \emph{spanning coresets for $\ell_p$ subspace approximation} with nearly optimal size. The $\ell_p$ subspace approximation is a popular generalization of the classic Frobenius norm low rank approximation problem, where the input is a set of $n$ points $\{\bfa_i\}_{i=1}^n$ in $d$ dimensions, and we wish to compute a rank $k$ subspace $F\subseteq\mathbb R^d$ that minimizes
\[
    \sum_{i=1}^n \norm{\bfa_i^\top(\bfI_d - \bfP_F)}_2^p
\]
where $\bfP_F$ denotes the orthogonal projection matrix onto $F$. Equivalently, we can write this as
\[
    \min_{\rank(F) \leq k}\norm{\bfA(\bfI_d - \bfP_F)}_{p,2}^p.
\]

While strong and weak coresets for this problem have attracted much attention \cite{FL2011, SV2012, SW2018, HV2020, FKW2021, WY2023b}, our main contribution to this line of research is on a different coreset guarantee, which we call \emph{spanning coresets}. Spanning coresets are subsets of the points $\bfa_i$ which span a $(1+\eps)$-optimal rank $k$ subspace, and is another popular guarantee in this literature \cite{DV2007, SV2012, CW2015a}. In addition to being an interesting object in its own right \cite{SV2012}, the existence of small spanning coresets have found applications to constructions for strong and weak coresets for $\ell_p$ subspace approximation \cite{HV2020}.

\begin{Definition}[Spanning coreset]
Let $\{\bfa_i\}_{i=1}^n\subseteq\mathbb R^d$. A subset $S\subseteq[n]$ is a \emph{$(1+\eps)$-spanning coreset} if the points $\{\bfa_i\}_{i\in S}$ span a $k$-dimensional subspace $\hat F$ such that
\[
    \norm{\bfA(\bfI_d - \bfP_{\hat F})}_{p,2}^p \leq (1+\eps)\min_{\rank(F) \leq k}\norm{\bfA(\bfI_d - \bfP_F)}_{p,2}^p.
\]
\end{Definition}

Our main result is the following upper bound on the size of spanning coresets. 

\begin{restatable}{Theorem}{SpanningCoreset}
\label{thm:spanning-coreset}
Let $\{\bfa_i\}_{i=1}^n\subseteq\mathbb R^d$, $1\leq p < \infty$, $k\in\mathbb N$, and $0<\eps<1$. Then, there exists a $(1+\eps)$-spanning coreset $S$ of size at most
\[
    \abs{S} = \begin{cases}
        O(\eps^{-2}k)(\log(k/\eps))^3 & p = 1 \\
        O(\eps^{-1}k)(\log(k/\eps))^3 & 1 < p \leq 2 \\
        O(\eps^{1-p}k^{p/2})(\log(k/\eps))^3 & 2 < p < \infty \\
    \end{cases}
\]
\end{restatable}

In particular, we improve the previous best result of $O(\eps^{-1} k^2 \log(k/\eps))$ due to Theorem 3.1 of \cite{SV2012} in the $k$ dependence for all $1 \leq p < 4$. The proof of this result is given in Section \ref{sec:subspace-approx}. Furthermore, we give the first lower bounds on the size of spanning coresets by generalizing an argument of \cite{DV2006} for $p = 2$, showing that spanning coresets must have size at least $\Omega(\eps^{-1}k)$ in Theorem \ref{thm:rank-k-lower-bound}. Together, our results settle the size of spanning coresets up to polylogarithmic factors for $1 < p < 2$. To obtain this result, we again use Dvoretzky's theorem to embed the problem to an embedded entrywise $\ell_p$ norm problem, and then apply our weak coreset results.

Finally, we note that our spanning coreset lower bound implies other interesting lower bounds for coresets. First, we note that weak coresets for $\ell_p$ subspace approximation are automatically spanning coresets, so our lower bound for spanning coresets also gives the first nontrivial lower bound on the size of weak coresets for $\ell_p$ subspace approximation. Secondly, we note that our proof of Theorem \ref{thm:spanning-coreset} in fact shows that any upper bound on weak coresets for $\ell_p$ regression with an embedding implies upper bounds for spanning coresets of the same size. Thus, our spanning coreset lower bound in fact implies an $\Omega(d/\eps)$ lower bound on the size of weak coresets for $\ell_p$ regression with an embedding, which establishes that our weak coreset upper bound for $\ell_p$ regression (Theorem \ref{thm:weak-coreset-informal}) is also nearly optimal for $1 < p < 2$ up to polylogarithmic factors. 

On the other hand, for $p>2$, our weak coreset lower bound of Theorem \ref{thm:weak-coreset-lower-bound} shows that our technique of reducing spanning coresets to weak coresets cannot prove a better upper bound than the result of Theorem \ref{thm:spanning-coreset}, and thus new ideas are required to improve upon the $\tilde O(\eps^{-1}k^2)$ spanning coreset upper bound of Theorem 3.1 of  \cite{SV2012}. This is an interesting open problem.

\subsection{Open directions}

We conclude with several potential directions for future research. One interesting question is to improve our understanding of upper bounds and lower bounds for coresets for single response $\ell_p$ regression.

\begin{Question}
How many rows are necessary and sufficient for strong and weak coresets for single response $\ell_p$ regression? 
\end{Question}

For strong coresets, this questions is already nearly optimally settled for $p < 2$ with $\tilde\Theta(\eps^{-2}d)$ rows known to be necessary and sufficient \cite{LWW2021}. For $p>2$, however, there is still a gap in our understanding, with the best known upper bound being $\tilde\Theta(\eps^{-2}d^{p/2})$ via $\ell_p$ Lewis weight sampling while the best known lower bound is only $\Omega(\eps^{-1}d^{p/2} + \eps^{-2}d)$. It is an interesting question to determine whether the lower bound can be improved to match the $\ell_p$ Lewis weight sampling upper bound or not. 

For weak coresets, the deficiencies are much more glaring. There are currently no known nontrivial lower bounds for weak coresets, while the best known algorithms are the better of the two upper bounds given by active $\ell_p$ regression and strong coresets, both of which are substantially more restricted settings than weak coresets.

Finally, we highlight the question of obtaining a nearly optimal upper bound on spanning coresets for $\ell_p$ subspace approximation for $p>2$. 

\begin{Question}
How many rows are necessary and sufficient for spanning coresets for $\ell_p$ subspace approximation?
\end{Question}

We conjecture that our lower bound of $\Omega(k/\eps)$ is tight, while the best known upper bound is the better of our Theorem \ref{thm:spanning-coreset} and $\tilde O(\eps^{-1}k^2)$ \cite{SV2012}.

\section{Preliminaries}

\subsection{\texorpdfstring{$\ell_p$}{lp} Lewis weights}

\begin{Definition}[One-sided $\ell_p$ Lewis weights \cite{JLS2022, WY2022}]\label{def:one-sided-lewis}
Let $\bfA\in\mathbb R^{n\times d}$ and $p\in(0,\infty)$. Let $\gamma\in(0,1]$. Then, weights $\bfw\in\mathbb R^n$ are \emph{$\gamma$-one-sided $\ell_p$ Lewis weights} if $\bfw_i \geq \gamma \cdot \bftau_i(\bfW^{1/2-1/p}\bfA)$, where $\bfW\coloneqq\diag(\bfw)$. If $\gamma = 1$, we just say that $\bfw$ are \emph{one-sided $\ell_p$ Lewis weights}.
\end{Definition}

The following theorem collects the results of \cite{CP2015, JLS2022} on the fastest known algorithms for approximating one-sided $\ell_p$ Lewis weights:

\begin{Theorem}
Let $\bfA\in\mathbb R^{n\times d}$ and $p>0$. There is an algorithm which computes one-sided $\ell_p$ Lewis weights (Def.~\ref{def:one-sided-lewis}) $\bfw$ such that $d\leq \norm{\bfw}_1 \leq 2d$ in $\tilde O(\nnz(\bfA) + d^\omega)$ time.
\end{Theorem}

\section{Strong coresets}

\begin{Theorem}[Strong coresets for multiple $\ell_p$ regression]
\label{thm:strong-coreset}
Let $\hat\bfX\in\mathbb R^{d\times m}$ satisfy
\[
    \norm{\bfA\hat\bfX-\bfB}_{p,p}^p \leq O(1)\min_{\bfX\in\mathbb R^{d\times m}}\norm{\bfA\bfX-\bfB}_{p,p}^p
\]
and let $\hat\bfB \coloneqq \bfA\hat\bfX-\bfB$. Let $\bfS$ be the $\ell_p$ sampling matrix (Definition \ref{def:sampling-matrix}) with sampling probabilities $q_i \geq \min\{1, \bfw_i/\alpha + \bfv_i/\beta\}$ for $\gamma$-one-sided $\ell_p$ Lewis weights $\bfw\in\mathbb R^n$, $\bfv_i = \norm{\bfe_i^\top\hat\bfB}_p^p/\norm{\hat\bfB}_{p,p}^p$,
\[
    \alpha = \begin{dcases}
        O(\gamma) \eps^2\bracks*{(\log d)^2\log n + \log\frac{1}\delta}^{-1} & p < 2 \\
        \frac{O(\gamma^{p/2})\eps^p}{\norm{\bfw}_1^{p/2-1}}\bracks*{(\log d)^2\log n + \log\frac{1}\delta}^{-1} & p > 2
    \end{dcases}
\]
and $\beta = O(\eps^{-2}\log\frac1\delta)$. Then with probability at least $1-\delta$,
\[
    \norm{\bfS(\bfA\bfX-\bfB)}_{p,p}^p = (1\pm\eps)\norm{\bfA\bfX-\bfB}_{p,p}^p
\]
simultaneously for every $\bfX\in\mathbb R^{d\times m}$.
\end{Theorem}

Our main technical lemma is the following result which generalizes the sampling results of \cite{MMWY2022, WY2023b} on preserving differences. The proof can be found in Appendix \ref{sec:lewis-diff}.

\begin{restatable}{Theorem}{LewisDiff}
\label{thm:lewis-weight-sampling-diff-cor}
Let $\bfS$ be the $\ell_p$ sampling matrix (Definition \ref{def:sampling-matrix}) with sampling probabilities $q_i \geq \min\{1, \bfw_i/\alpha\}$ for $\gamma$-one-sided $\ell_p$ Lewis weights $\bfw\in\mathbb R^n$ and
\[
    \alpha = \begin{dcases}
        \frac{O(\gamma)\eps^2}{\eta^{2/p}}\bracks*{(\log d)^2\log n + \log\frac1\delta}^{-1} & p < 2 \\
        \frac{O(\gamma^{p/2})\eps^p}{\eta \norm{\bfw}_1^{p/2-1}}\bracks*{(\log d)^2\log n + \log\frac1\delta}^{-1} & p > 2
    \end{dcases}.
\]
For each $\bfx^*\in\mathbb R^d$ and $\bfb^* = \bfA\bfx^*-\bfb$, with probability at least $1-\delta$,
\begin{align*}
    \abs*{\parens*{\norm{\bfS(\bfA\bfx-\bfb)}_p^p - \norm{\bfS\bfb^*}_p^p} - \parens*{\norm{\bfA\bfx-\bfb}_p^p - \norm{\bfb^*}_p^p}} \leq \eps \parens*{\norm{\bfb^*}_p^p + \norm{\bfS\bfb^*}_p^p + \frac1\eta \norm{\bfA\bfx-\bfA\bfx^*}_p^p}
\end{align*}
simultaneously for every $\bfx\in\mathbb R^d$.
\end{restatable}

Given Theorem \ref{thm:lewis-weight-sampling-diff-cor}, the proof of Theorem \ref{thm:strong-coreset} proceeds as described in the introduction. 

\begin{proof}[Proof of Theorem \ref{thm:strong-coreset}]
By replacing $\bfB$ by $\hat\bfB - \bfA\hat\bfX$, we assume that $\norm{\bfB}_p = O(\OPT)$. We apply Theorem \ref{thm:lewis-weight-sampling-diff-cor} with failure probability at $\eps^p\delta^2$. Now let $S\subseteq[m]$ be the set of columns for which the guarantee of Theorem \ref{thm:lewis-weight-sampling-diff-cor} fails. Note then that by Markov's inequality,
\[
    \sum_{j\in S}\norm{\bfB\bfe_j}_p^p = O(\eps^p \delta)\norm{\bfB}_{p,p}^p
\]
with probability at least $1-\delta$. We also have that
\[
    \sum_{j\in S}\norm{\bfS\bfB\bfe_j}_p^p \leq \frac1\delta \sum_{j\in S}\norm{\bfB\bfe_j}_p^p = O(\eps^p)\norm{\bfB}_{p,p}^p
\]
with probability at least $1-\delta$, again by Markov's inequality. Then,
\begin{align*}
    \norm{\bfS(\bfA\bfX-\bfB)\bfe_j}_p^p &= (1\pm\eps)\norm{\bfS\bfA\bfX\bfe_j}_p^p \pm \frac{O(1)}{\eps^{p-1}}\norm{\bfS\bfB\bfe_j}_p^p \\
    &= (1\pm \eps)^2\norm{\bfA\bfX\bfe_j}_p^p \pm \frac{O(1)}{\eps^{p-1}}\norm{\bfS\bfB\bfe_j}_p^p
\end{align*}
by using that $\bfS$ is a subspace embedding. Similarly, we have that
\[
    \norm{(\bfA\bfX-\bfB)\bfe_j}_p^p = (1\pm \eps)\norm{\bfA\bfX\bfe_j}_p^p \pm \frac{O(1)}{\eps^{p-1}}\norm{\bfB\bfe_j}_p^p.
\]
Then summing over $j\in S$ gives that
\[
    \sum_{j\in S}\norm{\bfS(\bfA\bfX-\bfB)\bfe_j}_p^p = \sum_{j\in S}\norm{(\bfA\bfX-\bfB)\bfe_j}_p^p \pm O(\eps)\norm{\bfB}_{p,p}^p.
\]
On the other hand, for $j\notin S$, Theorem \ref{thm:lewis-weight-sampling-diff-cor} succeeds so we have
\begin{align*}
    \norm{\bfS(\bfA\bfX-\bfB)\bfe_j}_p^p &= \norm{(\bfA\bfX-\bfB)\bfe_j}_p^p - \norm{\bfB\bfe_j}_p^p + \norm{\bfS\bfB\bfe_j}_p^p \pm\eps \parens*{\norm{\bfB\bfe_j}_p^p + \norm{\bfS\bfB\bfe_j}_p^p + \norm{\bfA\bfX\bfe_j}_p^p}
\end{align*}

Summing the guarantee over the $m$ columns $j$ gives
\begin{align*}
    \norm{\bfS(\bfA\bfX-\bfB)}_{p,p}^p &= \norm{\bfA\bfX-\bfB}_{p,p}^p-\norm{\bfB}_{p,p}^p+\norm{\bfS\bfB}_{p,p}^p \pm O(\eps)\parens*{\norm{\bfB}_{p,p}^p + \norm{\bfA\bfX}_{p,p}^p} \\
    &= \norm{\bfA\bfX-\bfB}_{p,p}^p\pm\eps\norm{\bfB}_{p,p}^p \pm O(\eps)\parens*{\norm{\bfB}_{p,p}^p + \norm{\bfA\bfX}_{p,p}^p} \\
    &= \norm{\bfA\bfX-\bfB}_{p,p}^p \pm O(\eps) \norm{\bfA\bfX-\bfB}_{p,p}^p.\qedhere
\end{align*}
\end{proof}

\section{Weak coresets}

We sketch the proof of the following result in this section. Full proofs can be found in Appendix \ref{sec:weak-coreset-proofs}.

\begin{Theorem}[Weak coresets for multiple $\ell_p$ regression]
\label{thm:weak-coreset-multiple-regression}
Let $\bfS$ be the $\ell_p$ sampling matrix (Definition \ref{def:sampling-matrix}) with sampling probabilities $q_i \geq \min\{1, \bfw_i/\alpha\}$ for $\gamma$-one-sided $\ell_p$ Lewis weights $\bfw\in\mathbb R^n$ and
\[
    \alpha = O(\gamma) \eps \delta^2\bracks*{(\log d)^2\log n + \log\frac{1}\delta}^{-1}\bracks*{\log\log\frac1\eps}^{-2}
\]
for $p<2$ and
\[
    \alpha = 
        \frac{O(\gamma^{p/2})\eps^{p-1} \delta^p}{\norm{\bfw}_1^{p/2-1}}\bracks*{(\log d)^2\log n + \log\frac{1}\delta}^{-1}\bracks*{\log\log\frac1\eps}^{-p}
\]
for $p>2$.
Then, for any $\hat\bfX\in\mathbb R^{d\times t}$ such that
\[
    \norm{\bfS(\bfA\hat\bfX\bfG-\bfB)}_{p,p}^p \leq (1+\eps)\min_{\bfX\in\mathbb R^{d\times t}} \norm{\bfS(\bfA\bfX\bfG-\bfB)}_{p,p}^p,
\]
we have
\[
    \norm{\bfA\hat\bfX\bfG-\bfB}_{p,p}^p \leq (1+O(\eps))\min_{\bfX\in\mathbb R^{d\times t}} \norm{\bfA\bfX\bfG-\bfB}_{p,p}^p.
\]
\end{Theorem}

We first establish lemmas that relate approximation quality to the closeness of solutions to the optimum in Section \ref{sec:closeness}, and we use this in an iterative argument in Section \ref{sec:iterative-reduction}.

\subsection{Closeness of nearly optimal solutions}
\label{sec:closeness}

The following lemma uses strong convexity for $p<2$ and a Bregman divergence bound for $p > 2$ to quantify the difference between the $\ell_p$ norms of two vectors.

\begin{Lemma}
\label{lem:closeness-vec}
For any $\bfy,\bfy'\in\mathbb R^n$, we have
\[
    \norm{\bfy'}_p^2 \geq \norm{\bfy}_p^2 - 2\norm{\bfy}_p^{2-p}\angle{\bfy^{\circ(p-1)}, \bfy-\bfy'} + \frac{p-1}{2}\norm{\bfy-\bfy'}_p^2
\]
if $1 < p < 2$ (Lemma 8.1 of \cite{BMN2001}) and
\[
    \norm{\bfy'}_p^p \geq \norm{\bfy}_p^p - p\angle{\bfy^{\circ(p-1)}, \bfy-\bfy'} + \frac{p-1}{p2^p}\norm{\bfy-\bfy'}_p^p
\]
if $2\leq p < \infty$ (Lemmas 3.2 and 4.6 of \cite{AKPS2019}).
\end{Lemma}

We need the following elementary computation.

\begin{Lemma}[Gradients of multiple $\ell_p$ regression]
\label{lem:grad-multiple-regression}
The gradient $\nabla_\bfX\norm*{\bfA\bfX\bfG - \bfB}_{p,p}^p$ is given by the formula
\[
    \sum_{i=1}^n \sum_{j=1}^m p[\bfA\bfX\bfG - \bfB](i, j)^{\circ(p-1)} (\bfA^\top\bfe_i)(\bfe_j^\top\bfG^\top)
\]
\end{Lemma}

The following lemma uses Lemmas \ref{lem:closeness-vec} and \ref{lem:grad-multiple-regression} to show that if $\bfX$ achieves a nearly optimal value, then $\bfX$ must be close to the optimal solution $\bfX^*$.

\begin{Lemma}[Closeness of nearly optimal solutions]
\label{lem:closeness}
Let $p>1$. For any $\bfX\in\mathbb R^{d\times t}$ such that $\norm*{\bfA\bfX\bfG-\bfB}_{p,p} \leq (1+\eta) \OPT$ with $\eta\in(0,1)$, we have that
\[
    \norm*{\bfA\bfX\bfG-\bfA\bfX^*\bfG}_{p,p} \leq \begin{dcases} 
        O(\eta^{1/2}) \OPT & p < 2 \\
        O(\eta^{1/p}) \OPT & p > 2
    \end{dcases}
\]
where $\bfX^* \coloneqq \arg\min_{\bfX\in\mathbb R^{d\times t}}\norm*{\bfA\bfX\bfG-\bfB}_{p,p}$.
\end{Lemma}

\subsection{Iterative size reduction argument}
\label{sec:iterative-reduction}

We now sketch the proof of Theorem \ref{thm:weak-coreset-multiple-regression}. 

We will need the following initial result to seed our iterative argument. Note that the dependence on $\eps$ is suboptimal by an $\eps$ factor for every $1 < p < \infty$.

\begin{Lemma}
\label{lem:initial-iteration}
Let $\bfS$ be the $\ell_p$ sampling matrix (Definition \ref{def:sampling-matrix}) with sampling probabilities $q_i \geq \min\{1, \bfw_i/\alpha\}$ for $\gamma$-one-sided $\ell_p$ Lewis weights $\bfw\in\mathbb R^n$ and
\[
    \alpha = O(\gamma) (\eps\delta)^2\bracks*{(\log d)^2\log n + \log\frac{1}\delta}^{-1}
\]
for $1\leq p < 2$ and
\[
    \alpha = \frac{O(\gamma^{p/2})(\eps\delta)^{p}}{\norm{\bfw}_1^{p/2-1}}\bracks*{(\log d)^2\log n + \log\frac{1}\delta}^{-1}
\]
for $2 < p < \infty$. Then, for any $\hat\bfX\in\mathbb R^{d\times t}$ such that
\[
    \norm{\bfS(\bfA\hat\bfX\bfG-\bfB)}_{p,p}^p \leq (1+\eps)\min_{\bfX\in\mathbb R^{d\times t}} \norm{\bfS(\bfA\bfX\bfG-\bfB)}_{p,p}^p,
\]
we have
\[
    \norm{\bfA\hat\bfX\bfG-\bfB}_{p,p}^p \leq (1+O(\eps))\min_{\bfX\in\mathbb R^{d\times t}} \norm{\bfA\bfX\bfG-\bfB}_{p,p}^p.
\]
\end{Lemma}

Starting from this initial solution bound of Lemma \ref{lem:initial-iteration}, we can proceed via an iterative argument similar to those of \cite{MMWY2022, WY2023b} which alternates between using a bound on the closeness of the solution to the optimal solution to improve the approximation (Theorem \ref{thm:lewis-weight-sampling-diff-cor}), and using a bound on the approximation to improve the closeness to the optimum (Lemma \ref{lem:closeness}). More specifically, we can show that for $1<p<2$, a bound of $C/\eps^\beta$ on the sample complexity implies that a bound of $C/\eps^{2\beta/(1+\beta)}$ is sufficient as well. Iterating this argument starting from $\beta = 2$ due to Lemma \ref{lem:initial-iteration} for $O(\log\log\frac1\eps)$ iterations yields the desired bound of $C/\eps$, as claimed. Similarly, for $p>2$, a bound of $C/\eps^\beta$ implies a bound of $C/\eps^{p\beta/(1+\beta)}$, which results in a final bound of $C/\eps^{p-1}$, as claimed. The full details can be found in Appendix \ref{sec:weak-coreset-proofs}.

\section{Lower bounds}
\label{sec:lower-bounds}

In this section, we complement our various upper bounds with matching lower bounds. In the interest of space, the proofs are given in Appendix \ref{sec:lower-bounds-proofs}.

\begin{Theorem}
\label{thm:strong-coreset-lower-bound}
Let $2 < p < \infty$ be fixed. Let $\eps\in(0, 1)$ be less than some sufficiently small constant. Then, a strong coreset $\bfS$ for multiple $\ell_p$ regression requires $\nnz(\bfS) = \Omega(\eps^{-p}d^{p/2})$ non-zero rows.
\end{Theorem}

\begin{Theorem}
\label{thm:weak-coreset-lower-bound}
Let $2 < p < \infty$ be fixed. Let $\eps\in(0, 1)$ be less than some sufficiently small constant. Then, a weak coreset $\bfS$ for multiple $\ell_p$ regression requires $\nnz(\bfS) = \Omega(\eps^{1-p}d^{p/2})$ non-zero rows.
\end{Theorem}

\begin{Theorem}
\label{thm:rank-k-lower-bound}
Let $1 \leq p < \infty$ and
\[
    c_p = \begin{cases}
        1/6 & p \leq 2 \\
        1/(6\cdot 5^{p/2-1}) & p > 2
    \end{cases}
\]
Let $k\in\mathbb N$. Then, there is a matrix $\bfB\in\mathbb R^{n\times (n+1)}$ such that for every $\eps \geq k/n$ and any subset of $s \leq (c_p/4) \eps^{-1} k$ rows, any rank $k$ subspace $F'$ spanned by the $s$ rows must have
\[
    \norm{\bfB\bfP_{F'} - \bfB}_{p,2}^p > (1+\eps) \min_{\rank(F) \leq k}\norm{\bfB\bfP_F - \bfB}_{p,2}^p.
\]
\end{Theorem}

\section*{Acknowledgements}

We thank the anonymous reviewers for useful feedback on improving the presentation of this work. David P.\ Woodruff and Taisuke Yasuda were supported by a Simons Investigator Award.

\bibliographystyle{alpha}
\bibliography{citations}

\newcommand{\etalchar}[1]{$^{#1}$}
\begin{thebibliography}{MMWY22}

\bibitem[AKPS19]{AKPS2019}
Deeksha Adil, Rasmus Kyng, Richard Peng, and Sushant Sachdeva.
\newblock Iterative refinement for $\ell_p$-norm regression.
\newblock In Timothy~M. Chan, editor, {\em Proceedings of the Thirtieth Annual
  {ACM-SIAM} Symposium on Discrete Algorithms, {SODA} 2019, San Diego,
  California, USA, January 6-9, 2019}, pages 1405--1424. {SIAM}, 2019.

\bibitem[BLM89]{BLM1989}
J.~Bourgain, J.~Lindenstrauss, and V.~Milman.
\newblock Approximation of zonoids by zonotopes.
\newblock {\em Acta Math.}, 162(1-2):73--141, 1989.

\bibitem[BMN01]{BMN2001}
Aharon Ben{-}Tal, Tamar Margalit, and Arkadi Nemirovski.
\newblock The ordered subsets mirror descent optimization method with
  applications to tomography.
\newblock {\em SIAM J. Optim.}, 12(1):79--108, 2001.

\bibitem[CD21]{CD2021}
Xue Chen and Michal Derezinski.
\newblock Query complexity of least absolute deviation regression via robust
  uniform convergence.
\newblock In Mikhail Belkin and Samory Kpotufe, editors, {\em Conference on
  Learning Theory, {COLT} 2021, 15-19 August 2021, Boulder, Colorado, {USA}},
  volume 134 of {\em Proceedings of Machine Learning Research}, pages
  1144--1179. {PMLR}, 2021.

\bibitem[Cla05]{Cla2005}
Kenneth~L. Clarkson.
\newblock Subgradient and sampling algorithms for $\ell_1$ regression.
\newblock In {\em Proceedings of the Sixteenth Annual ACM-SIAM Symposium on
  Discrete Algorithms}, SODA '05, pages 257--266, USA, 2005. Society for
  Industrial and Applied Mathematics.

\bibitem[CNW16]{CNW2016}
Michael~B. Cohen, Jelani Nelson, and David~P. Woodruff.
\newblock Optimal approximate matrix product in terms of stable rank.
\newblock In Ioannis Chatzigiannakis, Michael Mitzenmacher, Yuval Rabani, and
  Davide Sangiorgi, editors, {\em 43rd International Colloquium on Automata,
  Languages, and Programming, {ICALP} 2016, July 11-15, 2016, Rome, Italy},
  volume~55 of {\em LIPIcs}, pages 11:1--11:14. Schloss Dagstuhl -
  Leibniz-Zentrum f{\"{u}}r Informatik, 2016.

\bibitem[CP15]{CP2015}
Michael~B. Cohen and Richard Peng.
\newblock L\({}_{\mbox{p}}\) row sampling by lewis weights.
\newblock In Rocco~A. Servedio and Ronitt Rubinfeld, editors, {\em Proceedings
  of the Forty-Seventh Annual {ACM} on Symposium on Theory of Computing, {STOC}
  2015, Portland, OR, USA, June 14-17, 2015}, pages 183--192. {ACM}, 2015.

\bibitem[CP19]{CP2019}
Xue Chen and Eric Price.
\newblock Active regression via linear-sample sparsification.
\newblock In Alina Beygelzimer and Daniel Hsu, editors, {\em Conference on
  Learning Theory, {COLT} 2019, 25-28 June 2019, Phoenix, AZ, {USA}}, volume~99
  of {\em Proceedings of Machine Learning Research}, pages 663--695. {PMLR},
  2019.

\bibitem[CSS21]{CSS2021b}
Vincent Cohen{-}Addad, David Saulpic, and Chris Schwiegelshohn.
\newblock Improved coresets and sublinear algorithms for power means in
  euclidean spaces.
\newblock In Marc'Aurelio Ranzato, Alina Beygelzimer, Yann~N. Dauphin, Percy
  Liang, and Jennifer~Wortman Vaughan, editors, {\em Advances in Neural
  Information Processing Systems 34: Annual Conference on Neural Information
  Processing Systems 2021, NeurIPS 2021, December 6-14, 2021, virtual}, pages
  21085--21098, Virtual, 2021.

\bibitem[CW13]{CW2013}
Kenneth~L. Clarkson and David~P. Woodruff.
\newblock Low rank approximation and regression in input sparsity time.
\newblock In Dan Boneh, Tim Roughgarden, and Joan Feigenbaum, editors, {\em
  Symposium on Theory of Computing Conference, STOC'13, Palo Alto, CA, USA,
  June 1-4, 2013}, pages 81--90. {ACM}, 2013.

\bibitem[CW15]{CW2015a}
Kenneth~L. Clarkson and David~P. Woodruff.
\newblock Input sparsity and hardness for robust subspace approximation.
\newblock In Venkatesan Guruswami, editor, {\em {IEEE} 56th Annual Symposium on
  Foundations of Computer Science, {FOCS} 2015, Berkeley, CA, USA, 17-20
  October, 2015}, pages 310--329. {IEEE} Computer Society, 2015.

\bibitem[DDH{\etalchar{+}}09]{DDHKM2009}
Anirban Dasgupta, Petros Drineas, Boulos Harb, Ravi Kumar, and Michael~W.
  Mahoney.
\newblock Sampling algorithms and coresets for $\ell_p$ regression.
\newblock {\em {SIAM} J. Comput.}, 38(5):2060--2078, 2009.

\bibitem[DMM06a]{DMM2006}
Petros Drineas, Michael~W. Mahoney, and S.~Muthukrishnan.
\newblock Sampling algorithms for $\ell_2$ regression and applications.
\newblock In {\em Proceedings of the Seventeenth Annual {ACM-SIAM} Symposium on
  Discrete Algorithms, {SODA} 2006, Miami, Florida, USA, January 22-26, 2006},
  pages 1127--1136. {ACM} Press, 2006.

\bibitem[DMM06b]{DMM2006b}
Petros Drineas, Michael~W. Mahoney, and S.~Muthukrishnan.
\newblock Subspace sampling and relative-error matrix approximation:
  Column-row-based methods.
\newblock In Yossi Azar and Thomas Erlebach, editors, {\em Algorithms - {ESA}
  2006, 14th Annual European Symposium, Zurich, Switzerland, September 11-13,
  2006, Proceedings}, volume 4168 of {\em Lecture Notes in Computer Science},
  pages 304--314. Springer, 2006.

\bibitem[DV06]{DV2006}
Amit Deshpande and Santosh~S. Vempala.
\newblock Adaptive sampling and fast low-rank matrix approximation.
\newblock In Josep D{\'{\i}}az, Klaus Jansen, Jos{\'{e}} D.~P. Rolim, and Uri
  Zwick, editors, {\em Approximation, Randomization, and Combinatorial
  Optimization. Algorithms and Techniques, 9th International Workshop on
  Approximation Algorithms for Combinatorial Optimization Problems, {APPROX}
  2006 and 10th International Workshop on Randomization and Computation,
  {RANDOM} 2006, Barcelona, Spain, August 28-30 2006, Proceedings}, volume 4110
  of {\em Lecture Notes in Computer Science}, pages 292--303. Springer, 2006.

\bibitem[DV07]{DV2007}
Amit Deshpande and Kasturi~R. Varadarajan.
\newblock Sampling-based dimension reduction for subspace approximation.
\newblock In David~S. Johnson and Uriel Feige, editors, {\em Proceedings of the
  39th Annual {ACM} Symposium on Theory of Computing, San Diego, California,
  USA, June 11-13, 2007}, pages 641--650. {ACM}, 2007.

\bibitem[Dvo61]{Dvo1961}
Aryeh Dvoretzky.
\newblock Some results on convex bodies and {B}anach spaces.
\newblock In {\em Proc. {I}nternat. {S}ympos. {L}inear {S}paces ({J}erusalem,
  1960)}, pages 123--160. Jerusalem Academic Press, Jerusalem; Pergamon,
  Oxford, 1961.

\bibitem[FKW21]{FKW2021}
Zhili Feng, Praneeth Kacham, and David~P. Woodruff.
\newblock Dimensionality reduction for the sum-of-distances metric.
\newblock In Marina Meila and Tong Zhang, editors, {\em Proceedings of the 38th
  International Conference on Machine Learning, {ICML} 2021, 18-24 July 2021,
  Virtual Event}, volume 139 of {\em Proceedings of Machine Learning Research},
  pages 3220--3229. {PMLR}, 2021.

\bibitem[FL11]{FL2011}
Dan Feldman and Michael Langberg.
\newblock A unified framework for approximating and clustering data.
\newblock In Lance Fortnow and Salil~P. Vadhan, editors, {\em Proceedings of
  the 43rd {ACM} Symposium on Theory of Computing, {STOC} 2011, San Jose, CA,
  USA, 6-8 June 2011}, pages 569--578. {ACM}, 2011.

\bibitem[FLM77]{FLM1977}
T.~Figiel, J.~Lindenstrauss, and V.~D. Milman.
\newblock The dimension of almost spherical sections of convex bodies.
\newblock {\em Acta Math.}, 139(1-2):53--94, 1977.

\bibitem[HV20]{HV2020}
Lingxiao Huang and Nisheeth~K. Vishnoi.
\newblock Coresets for clustering in euclidean spaces: importance sampling is
  nearly optimal.
\newblock In Konstantin Makarychev, Yury Makarychev, Madhur Tulsiani, Gautam
  Kamath, and Julia Chuzhoy, editors, {\em Proccedings of the 52nd Annual {ACM}
  {SIGACT} Symposium on Theory of Computing, {STOC} 2020, Chicago, IL, USA,
  June 22-26, 2020}, pages 1416--1429. {ACM}, 2020.

\bibitem[JLS22]{JLS2022}
Arun Jambulapati, Yang~P. Liu, and Aaron Sidford.
\newblock Improved iteration complexities for overconstrained \emph{p}-norm
  regression.
\newblock In Stefano Leonardi and Anupam Gupta, editors, {\em {STOC} '22: 54th
  Annual {ACM} {SIGACT} Symposium on Theory of Computing, Rome, Italy, June 20
  - 24, 2022}, pages 529--542. {ACM}, 2022.

\bibitem[Lew78]{Lew1978}
D.~R. Lewis.
\newblock Finite dimensional subspaces of ${L}_p$.
\newblock {\em Studia Mathematica}, 63(2):207--212, 1978.

\bibitem[LLW23]{LLW2023}
Yi~Li, Honghao Lin, and David~P. Woodruff.
\newblock $\ell_p$-regression in the arbitrary partition model of
  communication.
\newblock In Gergely Neu and Lorenzo Rosasco, editors, {\em The Thirty Sixth
  Annual Conference on Learning Theory, {COLT} 2023, 12-15 July 2023,
  Bangalore, India}, volume 195 of {\em Proceedings of Machine Learning
  Research}, pages 4902--4928. {PMLR}, 2023.

\bibitem[LT91]{LT1991}
Michel Ledoux and Michel Talagrand.
\newblock {\em Probability in {B}anach spaces}.
\newblock Classics in Mathematics. Springer-Verlag, Berlin, 1991.
\newblock Isoperimetry and processes, Reprint of the 1991 edition.

\bibitem[LWW21]{LWW2021}
Yi~Li, Ruosong Wang, and David~P. Woodruff.
\newblock Tight bounds for the subspace sketch problem with applications.
\newblock {\em {SIAM} J. Comput.}, 50(4):1287--1335, 2021.

\bibitem[LWY21]{LWY2021}
Yi~Li, David~P. Woodruff, and Taisuke Yasuda.
\newblock Exponentially improved dimensionality reduction for $\ell_1$:
  Subspace embeddings and independence testing.
\newblock In Mikhail Belkin and Samory Kpotufe, editors, {\em Conference on
  Learning Theory, {COLT} 2021, 15-19 August 2021, Boulder, Colorado, {USA}},
  volume 134 of {\em Proceedings of Machine Learning Research}, pages
  3111--3195. {PMLR}, 2021.

\bibitem[Mil71]{Mil1971}
V.~D. Milman.
\newblock A new proof of {A}. {D}voretzky's theorem on cross-sections of convex
  bodies.
\newblock {\em Funkcional. Anal. i Prilo\v{z}en.}, 5(4):28--37, 1971.

\bibitem[MMR19]{MMR2019}
Konstantin Makarychev, Yury Makarychev, and Ilya~P. Razenshteyn.
\newblock Performance of johnson-lindenstrauss transform for \emph{k}-means and
  \emph{k}-medians clustering.
\newblock In Moses Charikar and Edith Cohen, editors, {\em Proceedings of the
  51st Annual {ACM} {SIGACT} Symposium on Theory of Computing, {STOC} 2019,
  Phoenix, AZ, USA, June 23-26, 2019}, pages 1027--1038. {ACM}, 2019.

\bibitem[MMWY22]{MMWY2022}
Cameron Musco, Christopher Musco, David~P. Woodruff, and Taisuke Yasuda.
\newblock Active linear regression for $\ell_p$ norms and beyond.
\newblock In {\em 63rd {IEEE} Annual Symposium on Foundations of Computer
  Science, {FOCS} 2022, Denver, CO, USA, October 31 - November 3, 2022}, pages
  744--753. {IEEE}, 2022.

\bibitem[PPP21]{PPP2021}
Aditya Parulekar, Advait Parulekar, and Eric Price.
\newblock {L1} regression with {L}ewis weights subsampling.
\newblock In Mary Wootters and Laura Sanit{\`{a}}, editors, {\em Approximation,
  Randomization, and Combinatorial Optimization. Algorithms and Techniques,
  {APPROX/RANDOM} 2021, August 16-18, 2021, University of Washington, Seattle,
  Washington, {USA} (Virtual Conference)}, volume 207 of {\em LIPIcs}, pages
  49:1--49:21. Schloss Dagstuhl - Leibniz-Zentrum f{\"{u}}r Informatik, 2021.

\bibitem[PTB13]{PTB2013}
Udaya Parampalli, Xiaohu Tang, and Serdar Boztas.
\newblock On the construction of binary sequence families with low correlation
  and large sizes.
\newblock {\em {IEEE} Trans. Inf. Theory}, 59(2):1082--1089, 2013.

\bibitem[PVZ17]{PVZ2017}
Grigoris Paouris, Petros Valettas, and Joel Zinn.
\newblock Random version of {D}voretzky's theorem in {$\ell_p^n$}.
\newblock {\em Stochastic Process. Appl.}, 127(10):3187--3227, 2017.

\bibitem[SV12]{SV2012}
Nariankadu~D. Shyamalkumar and Kasturi~R. Varadarajan.
\newblock Efficient subspace approximation algorithms.
\newblock {\em Discret. Comput. Geom.}, 47(1):44--63, 2012.

\bibitem[SW18]{SW2018}
Christian Sohler and David~P. Woodruff.
\newblock Strong coresets for k-median and subspace approximation: Goodbye
  dimension.
\newblock In Mikkel Thorup, editor, {\em 59th {IEEE} Annual Symposium on
  Foundations of Computer Science, {FOCS} 2018, Paris, France, October 7-9,
  2018}, pages 802--813. {IEEE} Computer Society, 2018.

\bibitem[SZ01]{SZ2001}
Gideon Schechtman and Artem Zvavitch.
\newblock Embedding subspaces of $l_p$ into $l_p^n$, $0<p<1$.
\newblock {\em Mathematische Nachrichten}, 227(1):133--142, 2001.

\bibitem[Tal90]{Tal1990}
Michel Talagrand.
\newblock Embedding subspaces of {$L_1$} into {$l^N_1$}.
\newblock {\em Proc. Amer. Math. Soc.}, 108(2):363--369, 1990.

\bibitem[Tal95]{Tal1995}
Michel Talagrand.
\newblock Embedding subspaces of {$L_p$} in {$l^N_p$}.
\newblock In {\em Geometric aspects of functional analysis ({I}srael,
  1992--1994)}, volume~77 of {\em Oper. Theory Adv. Appl.}, pages 311--325.
  Birkh\"{a}user, Basel, 1995.

\bibitem[Ver18]{Ver2018}
Roman Vershynin.
\newblock {\em High-dimensional probability}, volume~47 of {\em Cambridge
  Series in Statistical and Probabilistic Mathematics}.
\newblock Cambridge University Press, Cambridge, 2018.

\bibitem[WW19]{WW2019}
Ruosong Wang and David~P. Woodruff.
\newblock Tight bounds for $\ell_p$ oblivious subspace embeddings.
\newblock In Timothy~M. Chan, editor, {\em Proceedings of the Thirtieth Annual
  {ACM-SIAM} Symposium on Discrete Algorithms, {SODA} 2019, San Diego,
  California, USA, January 6-9, 2019}, pages 1825--1843. {SIAM}, 2019.

\bibitem[WY22]{WY2022}
David~P. Woodruff and Taisuke Yasuda.
\newblock High-dimensional geometric streaming in polynomial space.
\newblock In {\em 63rd {IEEE} Annual Symposium on Foundations of Computer
  Science, {FOCS} 2022, Denver, CO, USA, October 31 - November 3, 2022}, pages
  732--743. {IEEE}, 2022.

\bibitem[WY23a]{WY2023b}
David~P. Woodruff and Taisuke Yasuda.
\newblock New subset selection algorithms for low rank approximation: Offline
  and online.
\newblock In Barna Saha and Rocco~A. Servedio, editors, {\em Proceedings of the
  55th Annual {ACM} Symposium on Theory of Computing, {STOC} 2023, Orlando, FL,
  USA, June 20-23, 2023}, pages 1802--1813. {ACM}, 2023.

\bibitem[WY23b]{WY2023a}
David~P. Woodruff and Taisuke Yasuda.
\newblock Online {L}ewis weight sampling.
\newblock In Nikhil Bansal and Viswanath Nagarajan, editors, {\em Proceedings
  of the 2023 {ACM-SIAM} Symposium on Discrete Algorithms, {SODA} 2023,
  Florence, Italy, January 22-25, 2023}, pages 4622--4666. {SIAM}, 2023.

\bibitem[WY23c]{WY2023c}
David~P. Woodruff and Taisuke Yasuda.
\newblock Sharper bounds for $\ell_p$ sensitivity sampling.
\newblock In Andreas Krause, Emma Brunskill, Kyunghyun Cho, Barbara Engelhardt,
  Sivan Sabato, and Jonathan Scarlett, editors, {\em International Conference
  on Machine Learning, {ICML} 2023, 23-29 July 2023, Honolulu, Hawaii, {USA}},
  volume 202 of {\em Proceedings of Machine Learning Research}, pages
  37238--37272. {PMLR}, 2023.

\end{thebibliography}

\appendix

\section{\texorpdfstring{$\ell_p$}{lp} Lewis weight sampling for differences}
\label{sec:lewis-diff}

Throughout this section, we fix the following notation:

\begin{Definition}
\label{def:variables}
\leavevmode
\begin{itemize}
\item Let $1\leq p < \infty$. 
\item Let $\eps\in(0,1)$ be an accuracy parameter and let $\delta\in(0,1)$ be a failure probability parameter.
\item Let $\bfA\in\mathbb R^{n\times d}$ and $\bfb\in\mathbb R^n$. 
\item Let $\bfw\in\mathbb R^n$ be $\gamma$-one-sided $\ell_p$ Lewis weights for $\bfA$ such that $\max_{i=1}^n \bfw_i \leq w$. 
\item Let $\bfx^*\in\mathbb R^d$ any center, let $\eta\in(0,1)$ be a proximity parameter, and let $R\geq \norm{\bfA\bfx^*-\bfb}_p^p$ be a scale parameter. 
\item For each $i\in[n]$ and $\bfx\in\mathbb R^d$, let
\[
    \Delta_i(\bfx) \coloneqq \abs{[\bfA\bfx-\bfb](i)}^p - \abs{[\bfA\bfx^*-\bfb](i)}^p
\]
\end{itemize}
\end{Definition}

Our main result of the section is the following:

\begin{Theorem}
\label{thm:lewis-weight-sampling-diff}
Let $\bfS$ be the $\ell_p$ sampling matrix (Definition \ref{def:sampling-matrix}) with sampling probabilities $q_i \geq \min\{1, \bfw_i/\alpha\}$ for $\gamma$-one-sided $\ell_p$ Lewis weights $\bfw\in\mathbb R^n$ and
\[
    \alpha = \begin{dcases}
        O(\gamma) \frac{\eps^2}{\eta^{2/p}}\bracks*{(\log d)^2\log n + \log\frac1\delta}^{-1} & p < 2 \\
        O(\gamma^{p/2})\frac{\eps^p}{\eta \norm{\bfw}_1^{p/2-1}}\bracks*{(\log d)^2\log n + \log\frac1\delta}^{-1} & p > 2
    \end{dcases}.
\]
Then for each $\bfx^*\in\mathbb R^d$ and $R\geq \norm{\bfA\bfx^*-\bfb}_p^p$, with probability at least $1-\delta$,
\[
    \sup_{\norm{\bfA\bfx-\bfA\bfx^*}_p^p \leq \eta R}\abs*{\parens*{\norm{\bfS(\bfA\bfx-\bfb)}_p^p - \norm{\bfS(\bfA\bfx^*-\bfb)}_p^p} - \parens*{\norm{\bfA\bfx-\bfb}_p^p - \norm{\bfA\bfx^*-\bfb}_p^p}} \leq \eps (R + \norm{\bfS(\bfA\bfx^*-\bfb)}_p^p)
\]
\end{Theorem}

We will prove Theorem \ref{thm:lewis-weight-sampling-diff} throughout this section. Before doing so, we state the following more convenient form of the result:

\LewisDiff*
\begin{proof}
We apply Theorem \ref{thm:lewis-weight-sampling-diff} with $\delta$ set to $\delta/L$ for $L = O(\log(1/\delta\eps))$ and $R$ set to $2^l \norm{\bfA\bfx^*-\bfb}_p^p$ for $l\in[L]$. By a union bound, the conclusion holds simultaneously for every $l\in[L]$ with probability at least $1-\delta$. Furthermore, by Markov's inequality, $\norm{\bfS(\bfA\bfx^*-\bfb)}_p^p = O(1/\delta) \norm{\bfA\bfx^*-\bfb}_p^p$ with probability at least $1-\delta$.

If $\norm{\bfA\bfx-\bfA\bfx^*}_p^p \leq 2^L \norm{\bfA\bfx^*-\bfb}_p^p = \poly(1/\delta\eps) \norm{\bfA\bfx^*-\bfb}_p^p$, then the result follows immediately from applying the conclusion of Theorem \ref{thm:lewis-weight-sampling-diff} at the appropriate scale $l\in[L]$. Otherwise, we have that $\norm{\bfA\bfx-\bfA\bfx^*}_p^p \geq \poly(1/\delta\eps) \norm{\bfA\bfx^*-\bfb}_p^p$, in which case
\[
    \norm{\bfS(\bfA\bfx-\bfA\bfx^*)}_p^p \geq \Omega(1)\norm{\bfA\bfx-\bfA\bfx^*}_p^p \geq \poly(1/\delta\eps) \norm{\bfA\bfx^*-\bfb}_p^p
\]
so
\begin{align*}
    \norm*{\bfS(\bfA\bfx-\bfb)}_p^p - \norm*{\bfS(\bfA\bfx^*-\bfb)}_p^p &= (1\pm\eps)\norm*{\bfS(\bfA\bfx-\bfA\bfx^*)}_p^p \pm \frac{(1+\eps)^{p-1}}{\eps^{p-1}}\norm*{\bfS(\bfA\bfx^*-\bfb)}_p^p \\
    &= (1\pm\eps)\norm*{\bfS(\bfA\bfx-\bfA\bfx^*)}_p^p \pm \frac{(1+\eps)^{p-1}}{\delta\eps^{p-1}}\norm*{\bfA\bfx^*-\bfb}_p^p \\
    &= (1\pm O(\eps))\norm*{\bfS(\bfA\bfx-\bfA\bfx^*)}_p^p
\end{align*}
and similarly,
\[
    \norm*{\bfA\bfx-\bfb}_p^p - \norm*{\bfA\bfx^*-\bfb}_p^p = (1\pm O(\eps))\norm*{\bfA\bfx-\bfA\bfx^*}_p^p.
\]
Thus it suffices to have that
\[
    \abs*{\norm*{\bfS(\bfA\bfx-\bfA\bfx^*)}_p^p - \norm*{\bfA\bfx-\bfA\bfx^*}_p^p} \leq \frac{\eps}\eta \norm*{\bfA\bfx-\bfA\bfx^*}_p^p.
\]
In fact, standard $\ell_p$ Lewis weight sampling guarantees give
\[
    \abs*{\norm*{\bfS(\bfA\bfx-\bfA\bfx^*)}_p^p - \norm*{\bfA\bfx-\bfA\bfx^*}_p^p} \leq \begin{dcases}
        \frac{\eps}{\eta^{1/p}} \norm*{\bfA\bfx-\bfA\bfx^*}_p^p & p < 2 \\
        \frac{\eps^{p/2}}{\eta^{1/2}} \norm*{\bfA\bfx-\bfA\bfx^*}_p^p & p > 2
    \end{dcases}
\]
which is stronger.
\end{proof}

Throughout our proof of Theorem \ref{thm:lewis-weight-sampling-diff}, we will assume without loss of generality that $\bfS_{i,i}^p > 1$, that is we only consider rows that are sampled with probability $q_i < 1$, since rows that are kept with probability $q_i = 1$ do not contribute towards the sampling error. Note first that we can write
\[
    \abs*{\parens*{\norm{\bfS(\bfA\bfx-\bfb)}_p^p - \norm{\bfS(\bfA\bfx^*-\bfb)}_p^p} - \parens*{\norm{\bfA\bfx-\bfb}_p^p - \norm{\bfA\bfx^*-\bfb}_p^p}} = \abs*{\sum_{i=1}^n (\bfS_{i,i}^p-1)\Delta_i(\bfx)}.
\]
The supremum of this quantity, normalized by $(R + \norm{\bfS(\bfA\bfx^*-\bfb)}_p^p)^l$, over $\braces*{\norm{\bfA\bfx-\bfA\bfx^*}_p^p \leq \eta R}$ is a random variable. We will bound the $l$-th moment of this random variable for $l = O(\log\frac1\delta + \log n)$.

We start with a standard symmetrization procedure (see, e.g., \cite{CP2015, CD2021}).

\begin{Lemma}[Symmetrization]
\label{lem:general-symmetrization}
\begin{align*}
    &\E_{\bfS}\bracks*{\frac1{(R + \norm{\bfS(\bfA\bfx^*-\bfb)}_p^p)^l}\sup_{\norm{\bfA\bfx-\bfA\bfx^*}_p^p \leq \eta R}\abs*{\sum_{i=1}^n (\bfS_{i,i}^p-1) \Delta_i(\bfx)}^l} \\
    \leq~& 2^{l}\E_{\bfeps\sim\{\pm1\}^n,\bfS}\bracks*{\frac1{(R + \norm{\bfS(\bfA\bfx^*-\bfb)}_p^p)^l}\sup_{\norm{\bfA\bfx-\bfA\bfx^*}_p^p \leq \eta R}\abs*{\sum_{i=1}^n \bfeps_i \bfS_{i,i}^p \Delta_i(\bfx)}^l}
\end{align*}
\end{Lemma}

Next, we replace the Rademacher process on the right hand side of Lemma \ref{lem:general-symmetrization} by one which ``removes'' $\bfS_{i,i}^p$, that is, one of the form
\begin{equation}\label{eq:rp-to-bound}
    \E_{\bfeps\sim\{\pm1\}^n}\bracks*{\sup_{\norm{\bfA\bfx-\bfA\bfx^*}_p^p \leq \eta R}\abs*{\sum_{i=1}^n \bfeps_i \Delta_i(\bfx)}^l}.
\end{equation}
This is roughly done by noting that if we take $\bfS\bfA$ to be a ``part of'' $\bfA$, then the domain $\braces*{\norm{\bfA\bfx-\bfA\bfx^*}_p^p\leq\eta R}$ only dilates by a constant factor as $\bfS$ preserves $\ell_p$ norms in the column space of $\bfA$. More formally, we have the following lemma:

\begin{Lemma}
\label{lem:regularize-gp-active}
Let $\bfB\in\mathbb R^{m\times d}$ satisfy $\norm{\bfB\bfx}_p^p \leq C\norm{\bfA\bfx}_p^p$ for every $\bfx\in\mathbb R^d$. For every fixing of $\bfS$, let
\[
    \bfB_\bfS \coloneqq \begin{pmatrix}\bfS\bfA \\ \bfB\end{pmatrix}
\]
be the concatenation of $\bfS\bfA$ and $\bfB$, and let
\[
    F_\bfS = \sup_{\norm{\bfA\bfx}_p^p \leq 1} \abs*{\norm{\bfS\bfA\bfx}_p^p - \norm{\bfA\bfx}_p^p}.
\]
Suppose that for every fixing of $\bfS$ and $R'\geq R + \norm{\bfS(\bfA\bfx^*-\bfb)}_p^p$, we have that
\[
    \E_{\bfeps\sim\{\pm1\}^n} \sup_{\norm{\bfB_\bfS\bfx - \bfB_\bfS\bfx^*}_p^p\leq \eta R'}\abs*{\sum_{i=1}^{n} \bfeps_i \bfS_{i,i}^p\Delta_i(\bfx)} \leq \eps^l \delta R'^l 
\]
Then,
\[
    \E_{\bfS}\frac1{(R + \norm{\bfS(\bfA\bfx^*-\bfb)}_p^p)^l}\E_{\bfeps\sim\{\pm1\}^n}\sup_{\norm{\bfA\bfx-\bfA\bfx^*}_p^p \leq \eta R}\abs*{\sum_{i=1}^n \bfeps_i \bfS_{i,i}^p\Delta_i(\bfx)}^l \leq (2\eps)^l \delta \parens*{(1+C)^l + \E_\bfS[F_\bfS^l]}
\]
\end{Lemma}
\begin{proof}
Note that
\[
    \norm{\bfB_\bfS(\bfx - \bfx^*)}_p^p = \norm*{\bfS\bfA(\bfx-\bfx^*)}_p^p + \norm*{\bfB(\bfx-\bfx^*)}_p^p \leq (1 + F_\bfS + C)\norm{\bfA(\bfx-\bfx^*)}_p^p
\]
so
\begin{align*}
    \E_{\bfeps\sim\{\pm1\}^n}\sup_{\norm{\bfA\bfx-\bfA\bfx^*}_p^p \leq \eta R}\abs*{\sum_{i=1}^n \bfeps_i \bfS_{i,i}^p \Delta_i(\bfx)}^l &\leq \E_{\bfeps\sim\{\pm1\}^n}\sup_{\norm{\bfB_\bfS\bfx-\bfB_\bfS\bfx^*}_p^p \leq (1+F_\bfS+C)\eta R}\abs*{\sum_{i=1}^n \bfeps_i \bfS_{i,i}^p \Delta_i(\bfx)}^l \\
    &\leq \eps^l \delta (1+F_\bfS+C)^l (R+\norm{\bfS(\bfA\bfx^*-\bfb)}_p^p)^l \\
    &\leq \eps^l \delta 2^{l-1}((1+C)^l + F_\bfS^l) (R+\norm{\bfS(\bfA\bfx^*-\bfb)}_p^p)^l && \text{Fact \ref{fact:tri}}
\end{align*}
Taking expectations on both sides proves the lemma.
\end{proof}

Note that if $\bfS$ is the $\ell_p$ Lewis weight sampling matrix, then $\E[\abs{F_\bfS}^l]$ in Lemma \ref{lem:regularize-gp-active} is known to be bounded by $O(1)^l$ (that is, $\bfS$ is an $O(1)$-approximate $\ell_p$ subspace embedding) by standard results on $\ell_p$ Lewis weight sampling \cite{CP2015, WY2023a}.

Furthermore, we can design $\bfB$ such that the $\ell_p$ Lewis weights of $\bfB_\bfS$ are uniformly bounded by $\alpha$, where $\alpha$ is the oversampling parameter such that $\bfS$ samples the $i$th row with probability $\min\{1,\bfw_i/\alpha\}$. For $p<2$, this simply follows by taking $\bfB$ to be a flattening of $\bfA$ where every row is duplicated $1/\alpha$ times due to the monotonicity of $\ell_p$ Lewis weights \cite{CP2015}. For $p>2$, monotonicity of $\ell_p$ Lewis weights does not hold, but Theorem 5.2 of \cite{WY2023a} nonetheless shows that $\gamma$-one-sided $\ell_p$ Lewis weights can be constructed for $\bfB_\bfS$ with $\gamma = \Omega(1)$ that makes a similar argument go through.

Finally, it remains to bound the Rademacher process of the form of \eqref{eq:rp-to-bound}, where $\bfA$ has $\gamma$-one-sided $\ell_p$ Lewis weights uniformly bounded by $w = \alpha$. We will prove the following in Section \ref{sec:close-points}. Assuming this theorem, Theorem \ref{thm:lewis-weight-sampling-diff} follows by setting $w = \alpha$ as stated.

\begin{restatable}{Theorem}{ClosePoints}\label{thm:close-points}
For all $l\in\mathbb N$, we have
\begin{equation}\label{eq:diff-dist}
    \E_{\bfeps\sim\{\pm1\}^n}\sup_{\norm*{\bfA\bfx-\bfA\bfx^*}_p^p \leq \eta R}\abs*{\sum_{i=1}^n \bfeps_i\Delta_i(\bfx)}^l \leq \parens*{\eps R}^l
\end{equation}
where
\[
    \eps = \begin{dcases}
        O(w\eta^{2/p})^{1/2} \gamma^{-1/2}\bracks*{\parens*{(\log d)^2\log n}^{1+1/l} + l}^{1/2} & p < 2 \\
        O(w\eta \norm{\bfw}_1^{p/2-1})^{1/p}\gamma^{-1/2}\bracks*{\parens*{(\log d)^2\log n}^{1+1/l} + l}^{1/p} & p > 2
    \end{dcases}.
\]
\end{restatable}

\section{Rademacher process bounds}
\label{sec:close-points}

We continue to fix our notation from Definition \ref{def:variables}. We will prove Theorem \ref{thm:close-points} in this section.

We split the sum in \eqref{eq:diff-dist} into two parts: the part that is bounded by the $\gamma$-one-sided Lewis weights of $\bfA$, and the part that is not. To this end, define a threshold
\[
    \tau \coloneqq \begin{dcases}
        \frac{\eta}{\gamma^{p/2}\eps^{p}} & p < 2 \\
        \frac{\eta\norm*{\bfw}_1^{p/2-1}}{\gamma^{p/2}\eps^{p}} & p > 2
    \end{dcases}
\]
where $\eps$ will be determined later, and define the set of ``good'' entries $G\subseteq[n]$ as
\begin{equation}\label{eq:def-G}
    G \coloneqq \braces*{i\in[n] : \abs*{[\bfA\bfx^* - \bfb](i)} \leq \tau\bfw_i R}
\end{equation}
We then bound
\begin{align*}
    \E_{\bfeps\sim\{\pm1\}^n}\sup_{\norm*{\bfA\bfx-\bfA\bfx^*}_p^p \leq \eta R}\abs*{\sum_{i=1}^n \bfeps_i\Delta_i(\bfx)}^l &\leq 2^{l-1}\E_{\bfeps\sim\{\pm1\}^n}\sup_{\norm*{\bfA\bfx-\bfA\bfx^*}_p^p \leq \eta R}\abs*{\sum_{i\in G} \bfeps_i\Delta_i(\bfx)}^l \\
    &+ 2^{l-1}\E_{\bfeps\sim\{\pm1\}^n}\sup_{\norm*{\bfA\bfx-\bfA\bfx^*}_p^p \leq \eta R}\abs*{\sum_{i\in[n]\setminus G} \bfeps_i\Delta_i(\bfx)}^l
\end{align*}
using the Fact \ref{fact:tri}, and separately estimate each term. We can think of the first term as the ``sensitivity'' term, where each term in the sum is bounded by the Lewis weights of $\bfA$, and the latter term as the ``outlier'' term, where each term in the sum is much larger than the corresponding Lewis weights.

\subsection{Preliminaries}

We repeatedly use the following inequalities.

\begin{Fact}\label{fact:tri} 
For any $p \ge 1$ and any $a,b \in \mathbb R$, $|a+b|^p \le 2^{p-1} (|a|^p + |b|^p) = O(|a|^p + |b|^p)$.
\end{Fact}

\begin{Fact}[Corollary A.2, \cite{MMR2019}]\label{fact:relaxed-tri}
For any $p \ge 1$, $\eps>0$, and any $a,b \in \mathbb R$, $\abs{a+b}^p \leq (1+\eps)\abs{a}^p + \frac{(1+\eps)^{p-1}}{\eps^{p-1}}\abs{b}^p$.
\end{Fact}

\begin{Fact}\label{fact:p-conv} 
For any $p \ge 1$ and any $a,b \in \mathbb R$, $\abs{a}^p - \abs{b}^p \leq p\abs{a-b}(\abs{a}^{p-1} + \abs{b}^{p-1})$.
\end{Fact}

We will need the notion of weighted $\ell_p$ norms $\norm{\cdot}_{\bfw,p}$:

\begin{Definition}
Let $\bfw\in\mathbb R^n$ be non-negative weights. Then for $\bfy\in\mathbb R^n$, we define
\[
    \norm{\bfy}_{\bfw,p} \coloneqq \parens*{\sum_{i=1}^n \bfw_i \abs{\bfy(i)}^p}^{1/p}.
\]
\end{Definition}

\subsubsection{\texorpdfstring{$\ell_p$}{lp} Lewis weights}

\begin{Lemma}[One-sided Lewis weights bound sensitivities]\label{lem:one-sided-lewis-weights-bound-sensitivities}
Let $\bfA\in\mathbb R^{n\times d}$ and $0<p<\infty$. Let $\bfw\in\mathbb R^n$ be $\gamma$-one-sided $\ell_p$ Lewis weights. Then,
\[
    \sup_{\bfx\in\rowspan(\bfA)\setminus\{0\}} \frac{\abs*{[\bfA\bfx](i)}^p}{\norm*{\bfA\bfx}_p^p} \leq \begin{cases}
        \gamma^{-p/2}\norm*{\bfw}_1^{p/2-1}\cdot\bfw_i & p > 2 \\
        \gamma^{-1} \cdot \bfw_i & p < 2
    \end{cases}
\]
\end{Lemma}

\begin{Lemma}
\label{lem:lewis-lp-bounds-l2}
Let $\bfA\in\mathbb R^{n\times d}$ and let $\bfw$ be $\gamma$-one-sided $\ell_p$ Lewis weights for $\bfA$. Then,
\[
    \norm*{\bfW^{1/2-1/p}\bfA\bfx}_2 \leq \begin{cases}
        \norm*{\bfw}_1^{1/2-1/p}\norm*{\bfA\bfx}_p & p > 2 \\
        \gamma^{1/2-1/p}\norm{\bfA\bfx}_p & p < 2
    \end{cases}
\]
\end{Lemma}

\begin{Lemma}\label{lem:one-sided-lewis-basis-vs-weights}
Let $\bfA\in\mathbb R^{n\times d}$ and let $0<p<\infty$. The following hold:
Let $\bfw\in\mathbb R^n$ be $\gamma$-one-sided $\ell_p$ Lewis weights, and let $\bfR$ be a change of basis matrix $\bfR$ such that $\bfW^{1/2-1/p}\bfA\bfR$ is an orthonormal matrix. Then, for each $i\in[n]$,
\[
    \bfw_i \geq \gamma^{p/2}\cdot\norm*{\bfe_i^\top\bfA\bfR}_2^p.
\]
\end{Lemma}

\subsubsection{Gaussian processes}

\begin{Theorem}[Gaussian comparison, Equation 4.8, \cite{LT1991}]
\label{thm:gaussian-comparison}
Let $F:\mathbb R_+\to\mathbb R_+$ be convex and let $\{\bfx_i\}_{i=1}^n$ be a finite sequence in a Banach space. Then,
\[
    \E_{\bfeps\sim\{\pm1\}^n} F\parens*{\norm*{\sum_{i=1}^n \bfeps_i \bfx_i}} \leq \E_{\bfg\sim\mathcal N(0,\bfI_n)} F\parens*{\parens*{\frac{\pi}{2}}^{1/2}\norm*{\sum_{i=1}^n \bfg_i \bfx_i}}.
\]
\end{Theorem}

\begin{Theorem}[Dudley's entropy integral, Theorem 8.1.6, \cite{Ver2018}]\label{thm:dudley-tail}
Let $(X_t)_{t\in T}$ be a Gaussian process with pseudo-metric $d_X(s,t)\coloneqq \norm*{X_s - X_t}_2$. Let $E(T, d_X, u)$ denote the minimal number of $d_X$-balls of radius $u$ required to cover $T$. Then, for every $z\geq 0$, we have that
\[
	\Pr\braces*{\sup_{s, t\in T}\abs{X_s-X_t} \geq C\bracks*{\int_0^\infty \sqrt{\log E(T, d_X, u)}~du + z\cdot \diam(T)}} \leq 2\exp(-z^2)
\]
\end{Theorem}

Integrating the tail bound gives moment bounds. The following is taken from Lemma 6.8 of \cite{WY2023c}.

\begin{Lemma}[Moment bounds]
\label{lem:moment-bound}
Let $\Lambda$ be a Gaussian process with domain $T$ and distance $d_X$. Let $\mathcal E \coloneqq \int_0^\infty \sqrt{\log E(T, d_X, u)}~du$ and $\mathcal D = \diam(T)$. Then, for $l\in\mathbb N$,
\[
    \E_{\bfg\sim\mathcal N(0,\bfI_n)}[\abs{\Lambda}^l] \leq (2\mathcal E)^l (\mathcal E/\mathcal D) + O(\sqrt l \mathcal D)^{l}
\]
\end{Lemma}

\subsection{Estimates on the outlier term}

We first bound the outlier terms ($i\notin G$), which is much easier. 

\begin{Lemma}
\label{lem:outlier}
With probability $1$, we have that
\[
    \sup_{\norm*{\bfA\bfx-\bfA\bfx^*}_p^p \leq \eta R}\sum_{i\in[n]\setminus G}\abs*{\Delta_i(\bfx)} \leq O(\eps)R.
\]
\end{Lemma}
\begin{proof}
For each $i\in[n]\setminus G$, we have that
\begin{align*}
    \abs*{[\bfA\bfx-\bfb](i)} &\in \abs*{[\bfA\bfx^*-\bfb](i)} \pm \abs*{[\bfA\bfx^*-\bfA\bfx](i)} \\
    &\in \abs*{[\bfA\bfx^*-\bfb](i)} \pm \gamma^{-1/2}\norm*{\bfw}_1^{1/2-1/p}\bfw_i^{1/p}\norm*{\bfA\bfx^*-\bfA\bfx}_p && \text{Lemma \ref{lem:one-sided-lewis-weights-bound-sensitivities}} \\
    &\in \abs*{[\bfA\bfx^*-\bfb](i)} \pm \gamma^{-1/2}\eta^{1/p}\norm*{\bfw}_1^{1/2-1/p}\bfw_i^{1/p}R^{1/p} \\
    &\in \abs*{[\bfA\bfx^*-\bfb](i)} \pm \eps \abs*{[\bfA\bfx^*-\bfb](i)} && i\in[n]\setminus G
\end{align*}
Thus,
\[
    \abs*{\Delta_i(\bfx)} \leq O(\eps)\abs*{[\bfA\bfx^*-\bfb](i)}^p
\]
so
\[
    \sum_{i\in[n]\setminus G}\abs*{\Delta_i(\bfx)} \leq \sum_{i=1}^n O(\eps)\abs*{[\bfA\bfx^*-\bfb](i)}^p = O(\eps)\norm{\bfA\bfx^*-\bfb}_p^p \leq O(\eps)R.\qedhere
\]
\end{proof}

\subsection{Estimates on the sensitivity term}

Next, we estimate the sensitivity term ($i\in G$),
\[
    \E_{\bfeps\sim\{\pm1\}^n}\sup_{\norm*{\bfA\bfx-\bfA\bfx^*}_p^p \leq \eta R}\abs*{\sum_{i\in G} \bfeps_i\Delta_i(\bfx)}^l.
\]
To estimate this moment, we obtain a subgaussian tail bound via the tail form of Dudley's entropy integral, and then integrate it. We will crucially use that $\abs*{\Delta_i(\bfx)}$ for $i\in G$ is bounded over all $\norm*{\bfA\bfx-\bfA\bfx^*}_p^p \leq \eta R$, which gives the following sensitivity bound:

\begin{Lemma}
\label{lem:Delta-sensitivity-bound}
For all $i\in G$, and $\bfx\in\mathbb R^d$ with $\norm{\bfA\bfx-\bfA\bfx^*}_p^p \leq \eta R$, we have $\abs{[\bfA\bfx-\bfb](i)}^p \leq O(\tau \bfw_i R)$ and $\abs{\Delta_i(\bfx)}\leq O(\tau\bfw_i R)$.
\end{Lemma}
\begin{proof}
We have
\begin{align*}
    \abs*{[\bfA\bfx-\bfb](i)}^p &\leq 2^{p-1}\parens*{\abs*{[\bfA\bfx^*-\bfb](i)}^p + \abs*{[\bfA\bfx-\bfA\bfx^*](i)}^p} && \text{Fact \ref{fact:tri}} \\
    &\leq 2^{p-1}\tau\bfw_i R + 2^{p-1}\gamma^{-p/2}\eta\norm*{\bfw}_1^{0\lor (p/2-1)} \bfw_i R && \text{$i\in G$ (see \eqref{eq:def-G}) and Lemma \ref{lem:one-sided-lewis-weights-bound-sensitivities}} \\
    &\leq O(\tau\bfw_i R)
\end{align*}
The bound on $\Delta_i(\bfx)$ follows easily from the above calculation. 
\end{proof}

\subsubsection{Bounding low-sensitivity entries}

We now separately handle entries $i\in G$ with small Lewis weight. To do this end, define
\[
    J \coloneqq \braces*{i\in G : \bfw_i \geq \frac{\eps}{\tau n}}.
\]
We then bound the mass on the complement of $J$:

\begin{Lemma}\label{lem:not-J}
For all $\norm*{\bfA\bfx-\bfA\bfx^*}_p^p \leq \eta R$, we have that
\[
    \sum_{i\in [n]\setminus J} \abs*{\Delta_i(\bfx)} \leq O(\eps R)
\]
\end{Lemma}
\begin{proof}
We have that for each $i\in[n]\setminus J$, $\bfw_i \leq \eps/\tau n$ so by Lemma \ref{lem:Delta-sensitivity-bound},
\[
    \sum_{i\in [n]\setminus J} \abs*{\Delta_i(\bfx)} \leq \sum_{i\in[n]\setminus J} O(\tau \bfw_i R) \leq \sum_{i\in[n]\setminus J} \frac{O(\eps)}{n} R \leq O(\eps R)
\]
\end{proof}

\subsubsection{Bounding high-sensitivity entries: Gaussian processes}

Finally, it remains to bound the Rademacher process only on the entries indexed by $i\in J$. By a Gaussian comparison theorem (Theorem \ref{thm:gaussian-comparison}), we may bound the Rademacher process above by a Gaussian process instead, that is,
\begin{equation}
\label{eq:active-gp}
    \E_{\bfeps\sim\{\pm1\}^n}\sup_{\norm*{\bfA\bfx-\bfA\bfx^*}_p^p \leq \eta R}\abs*{\sum_{i\in J} \bfeps_i\Delta_i(\bfx)}^l \leq \parens*{\frac{\pi}{2}}^{l/2}\E_{\bfg\sim\mathcal N(0,\bfI_n)}\sup_{\norm*{\bfA\bfx-\bfA\bfx^*}_p^p \leq \eta R}\abs*{\sum_{i\in J} \bfg_i\Delta_i(\bfx)}^l.
\end{equation}
We can now appeal to the theory of Gaussian processes to bound this quantity. Define a Gaussian process by
\[
    X_\bfx \coloneqq \sum_{i\in J} \bfg_i\Delta_i(\bfx)
\]
with pseudo-metric
\[
    d_X(\bfx,\bfx') \coloneqq \parens*{\E_\bfg\abs*{X_\bfx-X_\bfx'}^2}^{1/2} = \parens*{\sum_{i\in J}(\Delta_i(\bfx)-\Delta_i(\bfx'))^2}^{1/2}
\]
We will use Dudley's entropy integral (Theorem \ref{thm:dudley-tail}) to bound the tail of this quantity, and then integrate to obtain moment bounds.

Using the sensitivity bound of Lemma \ref{lem:Delta-sensitivity-bound}, we obtain a bound on the pseudo-metric $d_X$. %

\begin{Lemma}
\label{lem:dX-bound-active}
Let $q = O(\log(\tau n/\eps))$. For $\bfx,\bfx'\in T$ for $T = \braces{\norm*{\bfA\bfx-\bfA\bfx^*}_p^p\leq \eta R}$, we have that
\[
    d_X(\bfx,\bfx') \leq \begin{dcases}
        O(w^{1/2}) \eta^{1/p-1/2}\norm{\bfW^{-1/p}\bfA(\bfx-\bfx')}_{\bfw,q}^{p/2} R^{1/2} & p < 2 \\
        O(w^{1/2}) \tau^{1/2-1/p}\norm{\bfW^{-1/p}\bfA(\bfx-\bfx')}_{\bfw,q} R^{1-1/p} & p > 2
    \end{dcases}
\]
and
\[
    \diam(T) = \sup_{\bfx,\bfx'\in T}d_X(\bfx,\bfx') \leq \begin{dcases}
        O(w^{1/2} \eta^{1/p}\gamma^{-1/2} R) & p < 2 \\
        O(\eps w^{1/2} \tau^{1/2} R) & p > 2
    \end{dcases}
\]
\end{Lemma}
\begin{proof}
Let $\bfy = \bfA\bfx - \bfb$ and $\bfy' = \bfA\bfx'-\bfb$. Note then that
\begin{align*}
    d_X(\bfx,\bfx')^2 &= \sum_{i\in J}(\Delta_i(\bfx) - \Delta_i(\bfx'))^2 = \sum_{i\in J}(\abs{\bfy(i)}^p - \abs{\bfy'(i)}^p)^2 \\
    &\leq p^2 \sum_{i\in J}  \abs{\bfy(i) - \bfy'(i)}^2(\abs{\bfy(i)}^{p-1} + \abs{\bfy'(i)}^{p-1})^2 && \text{Fact \ref{fact:p-conv}}
\end{align*}
For $p<2$, we have that
\begin{align*}
    d_X(\bfx,\bfx')^2 &\leq p^2 \norm{(\bfy-\bfy')\vert_J}_\infty^p\sum_{i\in J}  (\abs{\bfy(i) - \bfy'(i)})^{2-p}(\abs{\bfy(i)}^{p-1} + \abs{\bfy'(i)}^{p-1})^2 \\
    &\leq 2p^2 \norm{(\bfy-\bfy')\vert_J}_\infty^p \sum_{i\in J}  (\abs{\bfy(i) - \bfy'(i)})^{2-p}(\abs{\bfy(i)}^{2p-2} + \abs{\bfy'(i)}^{2p-2}) \\
    &\leq 2p^2 \norm{(\bfy-\bfy')\vert_J}_\infty^p \norm{\bfy-\bfy'}_p^{2-p} (\norm{\bfy}_p^{2p-2} + \norm{\bfy'}_p^{2p-2}) && \text{H\"older's inequality} \\
    &\leq O(\eta^{2/p-1})\norm*{(\bfy-\bfy')\vert_J}_\infty^p R.
\end{align*}
where H\"older's inequality is applied with exponents $\frac{p}{2-p} > 1$ and $\frac{p}{2p-2} > 1$. For $p>2$, we have that
\begin{align*}
    d_X(\bfx,\bfx')^2 &\leq 2p^2 \norm*{(\bfy-\bfy')\vert_J}_\infty^2\sum_{i=1}^n \abs{\bfy(i)}^{2p-2} + \abs{\bfy'(i)}^{2p-2} \\
    &\leq 2p^2 \max\{\norm*{\bfy\vert_J}_\infty,\norm{\bfy'\vert_J}_\infty\}^{p-2}\norm*{(\bfy-\bfy')\vert_J}_\infty^2\sum_{i=1}^n \abs{\bfy(i)}^{p} + \abs{\bfy'(i)}^{p} \\
    &\leq O(1) (\tau w R)^{1-2/p}\norm*{(\bfy-\bfy')\vert_J}_\infty^2 R && \text{Lemma \ref{lem:Delta-sensitivity-bound}}
\end{align*}
Furthermore, we have that
\begin{align*}
    \norm{(\bfy-\bfy')\vert_J}_\infty &= \norm*{(\bfA\bfx-\bfA\bfx')\vert_J}_\infty \\
    &= \norm{\bfW^{1/p}(\bfW^{-1/p}\bfA\bfx-\bfW^{-1/p}\bfA\bfx')\vert_J}_\infty \\
    &\leq w^{1/p}\norm{(\bfW^{-1/p}\bfA\bfx-\bfW^{-1/p}\bfA\bfx')\vert_J}_\infty \\
    &\leq 2w^{1/p}\norm{\bfW^{-1/p}\bfA\bfx-\bfW^{-1/p}\bfA\bfx'}_{\bfw,q}
\end{align*}
where the last step follows from the fact that $\bfw_i\geq \eps/\tau n$ for $i\in J$ and $q = O(\log(\tau n/\eps))$. Combining these bounds gives the claimed bound on $d_X(\bfx,\bfx')$.

Finally, we have by Lemma \ref{lem:one-sided-lewis-weights-bound-sensitivities} that
\[
    \norm{\bfW^{-1/p}\bfA(\bfx-\bfx^*)}_\infty = \max_{i=1}^n \frac{\abs{[\bfA(\bfx-\bfx^*)](i)}}{\bfw_i} \leq \begin{dcases}
        \gamma^{-1/p}\norm{\bfA(\bfx-\bfx^*)}_p & p < 2 \\
        \gamma^{-1/2}\norm{\bfw}_1^{1/2-1/p}\norm{\bfA(\bfx-\bfx^*)}_p & p > 2
    \end{dcases}
\]
so we have the claimed diameter bound for the set $\braces{\norm{\bfA(\bfx-\bfx^*)}_p^p \leq \eta R}$.
\end{proof}

The following entropy bounds are obtained from \cite{WY2023c}, which in turn largely follow \cite{BLM1989}.

\begin{Remark}
\label{rem:simple-net}
The following entropy bounds are net necessary if we only need this result for $d = 1$, for example for applications to Euclidean power means. In this case, standard volume arguments suffice (see, e.g., Lemma 2.4 of \cite{BLM1989}).
\end{Remark}

\begin{Lemma}\label{lem:p-2-q-cover}
Let $1\geq\bfw\in\mathbb R^n$ be non-negative weights. Let $2\leq q < \infty$ and let $\bfA\in\mathbb R^{n\times d}$ be such that $\bfW^{1/2}\bfA$ is orthonormal. Let $\tau \geq \max_{i=1}^n \norm*{\bfe_i^\top\bfA}_2^2$. Let $B^p_\bfw(\bfA) \coloneqq \braces{\bfx : \norm{\bfA\bfx}_{\bfw,p}\leq 1}$. Then,
\[
    \log E(B_\bfw^2(\bfA), B_{\bfw}^q(\bfA), t) \leq O(1) \frac{n^{2/q} q\cdot \tau}{t^2}
\]
and
\[
    \log E(B_\bfw^p(\bfA), B_{\bfw}^q(\bfA), t) \leq O(1) \frac1{t^p}\parens*{\frac{\log d}{2-p} + \log n + n^{2/q}q}\tau.
\]
for $p<2$.
\end{Lemma}

We may now evaluate Dudley's entropy integral.

\begin{Lemma}[Entropy integral bound for $p<2$]
\label{lem:active-lewis-entropy-int-p<2}
We have that
\[
    \int_0^\infty \sqrt{\log E(B^p(\bfA), d_{X}, t)}~dt \leq O(w^{1/2}\gamma^{-1/2}\eta^{1/2} R)\parens*{\log\frac{\tau n}{\eps}}^{1/2}\log d
\]
\end{Lemma}
\begin{proof}
Note that it suffices to integrate the entropy integral to $\diam(T)$, which is bounded in Lemma \ref{lem:dX-bound-active}. Note also that $T$ is just a translation of $(\eta R)^{1/p}\cdot B^p(\bfA)$, so we have
\begin{align*}
    \log E(T, d_X, t) &= \log E((\eta R)^{1/p}\cdot B^p(\bfA), d_X, t) \\
    &= \log E((\eta R)^{1/p}\cdot B^p(\bfA), K\norm{\bfW^{-1/p}\bfA(\cdot)}_{\bfw,q}^{p/2}, t) && \text{Lemma \ref{lem:dX-bound-active}} \\
    &= \log E(B_\bfw^p(\bfW^{-1/p}\bfA), B_\bfw^q(\bfW^{-1/p}\bfA), t^{2/p}/K^{2/p}(\eta R)^{1/p})
\end{align*}
where $K = O(w^{1/2} \eta^{1/p-1/2} R^{1/2})$. 

For small radii less than $\lambda$ for a parameter $\lambda$ to be chosen, we use a standard volume argument, which shows that
\[
    \log E(B_\bfw^p(\bfW^{-1/p}\bfA), B_\bfw^q(\bfW^{-1/p}\bfA), t) \leq O(d)\log\frac{n}{t}
\]
so
\begin{align*}
    \int_0^\lambda \sqrt{\log E(T, d_X, t)}~dt &\leq \int_0^\lambda \sqrt{d\log\frac{n K^{2/p}(\eta R)^{1/p}}{t^{2/p}}}~dt \\
    &\leq \lambda \sqrt{d\log(n (\eta^{2/p} w)^{1/p})} + \sqrt d\int_0^\lambda \sqrt{\log\frac{R^{2/p}}{t^{2/p}}}~dt \\
    &\leq \lambda \sqrt{d\log(n (\eta^{2/p} w)^{1/p})} + \sqrt d \cdot O(\lambda) \sqrt{\log\frac{R}\lambda} \\
    &\leq O(\lambda)\sqrt{d\log\frac{n (\eta^{2/p} w)^{1/p} R}{\lambda}}
\end{align*}
On the other hand, for large radii larger than $\lambda$, we use the bounds of Lemma \ref{lem:p-2-q-cover}. Note that the entropy bounds do not change if we replace $\bfA$ by $\bfA\bfR$, where $\bfR$ is the change of basis matrix such that $\bfW^{1/2-1/p}\bfA\bfR$ is orthonormal. Then by the properties of $\gamma$-one-sided $\ell_p$ Lewis weights (Lemma \ref{lem:one-sided-lewis-basis-vs-weights}), we have
\[
    \norm{\bfe_i^\top\bfW^{-1/p}\bfA\bfR}_2^2 = \bfw_i^{-2/p}\norm{\bfe_i^\top\bfA\bfR}_2^2 \leq \gamma^{-1}.
\]
Then, Lemma \ref{lem:p-2-q-cover} gives
\[
    \log E(B_\bfw^p(\bfW^{-1/p}\bfA), B_\bfw^q(\bfW^{-1/p}\bfA), t^{2/p}/K^{2/p}(\eta R)^{1/p}) = \frac{O(w\eta^{2/p} R^2)}{\gamma t^2}\log\frac{\tau n}{\eps}
\]
so the entropy integral gives a bound of
\begin{align*}
    \frac{O(w^{1/2}\eta^{1/p} R)}{\gamma^{1/2}}\parens*{\log\frac{\tau n}{\eps}}^{1/2}\int_\lambda^{\diam(T)} \frac1t~dt = \frac{O(w^{1/2}\eta^{1/p} R)}{\gamma^{1/2}}\parens*{\log\frac{\tau n}{\eps}}^{1/2}\log\frac{\diam(T)}{\lambda}.
\end{align*}
We choose $\lambda = \diam(T) / \sqrt{d}$, which yields the claimed conclusion.
\end{proof}

An analogous result and proof holds for $p>2$.

\begin{Lemma}[Entropy integral bound for $p>2$]
\label{lem:active-lewis-entropy-int-p>2}
Let $2<p<\infty$. Let $\bfA\in\mathbb R^{n\times d}$ and let $0\leq \bfw\in\mathbb R^n$ be $\gamma$-one-sided $\ell_p$ Lewis weights. Let $w = \max_{i\in [n]}\bfw_i$. Then,
\[
    \int_0^\infty \sqrt{\log E(B^p(\bfA), d_{X}, t)}~dt \leq O(w^{1/2}\eps \tau^{1/2} R)\parens*{\log\frac{\tau n}{\eps}}^{1/2}\log d
\]
\end{Lemma}
\begin{proof}
Note that it suffices to integrate the entropy integral to $\diam(T)$, which is bounded in Lemma \ref{lem:dX-bound-active}. Note also that $T$ is just a translation of $(\eta R)^{1/p}\cdot B^p(\bfA)$, so we have
\begin{align*}
    \log E(T, d_X, t) &= \log E((\eta R)^{1/p}\cdot B^p(\bfA), d_X, t) \\
    &= \log E((\eta R)^{1/p}\cdot B^p(\bfA), K\norm{\bfW^{-1/p}\bfA(\cdot)}_{\bfw,q}, t) && \text{Lemma \ref{lem:dX-bound-active}} \\
    &= \log E(B_\bfw^p(\bfW^{-1/p}\bfA), B_\bfw^q(\bfW^{-1/p}\bfA), t/K(\eta R)^{1/p})
\end{align*}
where $K = O(w^{1/2} \tau^{1/2-1/p} R^{1-1/p})$. 

For small radii less than $\lambda$ for a parameter $\lambda$ to be chosen, we use a standard volume argument, which shows that
\[
    \log E(B_\bfw^p(\bfW^{-1/p}\bfA), B_\bfw^q(\bfW^{-1/p}\bfA), t) \leq O(d)\log\frac{n}{t}
\]
so
\begin{align*}
    \int_0^\lambda \sqrt{\log E(T, d_X, t)}~dt &\leq \int_0^\lambda \sqrt{d\log\frac{n K(\eta R)^{1/p}}{t}}~dt \\
    &\leq \lambda \sqrt{d\log(n w^{1/2} \eta^{1/p} \tau^{1/2-1/p})} + \sqrt d\int_0^\lambda \sqrt{\log\frac{R}{t}}~dt \\
    &\leq \lambda \sqrt{d\log(n w^{1/2} \eta^{1/p} \tau^{1/2-1/p})} + \sqrt d \cdot O(\lambda) \sqrt{\log\frac{R}\lambda} \\
    &\leq O(\lambda)\sqrt{d\log\frac{n w^{1/2} \eta^{1/p} \tau^{1/2-1/p} R}{\lambda}}
\end{align*}
On the other hand, for large radii larger than $\lambda$, we use the bounds of Lemma \ref{lem:p-2-q-cover}. Note that the entropy bounds do not change if we replace $\bfA$ by $\bfA\bfR$, where $\bfR$ is the change of basis matrix such that $\bfW^{1/2-1/p}\bfA\bfR$ is orthonormal. Then by the properties of $\gamma$-one-sided $\ell_p$ Lewis weights (Lemma \ref{lem:one-sided-lewis-basis-vs-weights}), we have
\[
    \norm{\bfe_i^\top\bfW^{-1/p}\bfA\bfR}_2^2 = \bfw_i^{-2/p}\norm{\bfe_i^\top\bfA\bfR}_2^2 \leq \gamma^{-1}.
\]
Then, Lemma \ref{lem:lewis-lp-bounds-l2} and Lemma \ref{lem:p-2-q-cover} give
\begin{align*}
    &\log E(B_\bfw^p(\bfW^{-1/p}\bfA), B_\bfw^q(\bfW^{-1/p}\bfA), t/K(\eta R)^{1/p}) \\
    \leq~&\log E(B_\bfw^2(\bfW^{-1/p}\bfA), B_\bfw^q(\bfW^{-1/p}\bfA), t/K(\eta R)^{1/p}\norm{\bfw}_1^{1/2-1/p}) \\
    \leq~&\frac{K^2(\eta R)^{2/p} \norm{\bfw}_1^{1-2/p}}{\gamma t^2}\log\frac{\tau n}{\eps} \\
    \leq~&\frac{O(w) \eps^2 \tau R^2}{t^2}\log\frac{\tau n}{\eps}
\end{align*}
so the entropy integral gives a bound of
\begin{align*}
    O(w^{1/2}\eps \tau^{1/2} R)\parens*{\log\frac{\tau n}{\eps}}^{1/2}\int_\lambda^{\diam(T)} \frac1t~dt = O(w^{1/2}\eps \tau^{1/2} R)\parens*{\log\frac{\tau n}{\eps}}^{1/2}\log\frac{\diam(T)}{\lambda}.
\end{align*}
We choose $\lambda = \diam(T) / \sqrt{d}$, which yields the claimed conclusion.
\end{proof}

We are now ready to prove Theorem \ref{thm:close-points}.

\begin{proof}[Proof of Theorem \ref{thm:close-points}]
We have by Lemma \ref{lem:moment-bound} that the Gaussian process of \eqref{eq:active-gp} is bounded by
\[
    (2\mathcal E)^l (\mathcal E/\mathcal D) + O(\sqrt l \mathcal D)^l
\]
where
\[
    \mathcal E \leq \begin{dcases}
        O(w^{1/2}\gamma^{-1/2}\eta^{1/p} R)\parens*{\log\frac{\tau n}{\eps}}^{1/2}\log d & p < 2 \\
        O(\eps w^{1/2} \tau^{1/2} R)\parens*{\log\frac{\tau n}{\eps}}^{1/2}\log d & p > 2
    \end{dcases}
\]
by Lemmas \ref{lem:active-lewis-entropy-int-p<2} and \ref{lem:active-lewis-entropy-int-p>2} and
\[
    \mathcal D \leq \begin{dcases}
        O(w^{1/2} \eta^{1/p} \gamma^{-1/2} R) & p < 2 \\
        O(\eps w^{1/2} \tau^{1/2} R) & p > 2
    \end{dcases}
\]
by Lemma \ref{lem:dX-bound-active}. This gives a bound of $(\alpha R)^l$ on \eqref{eq:active-gp}, where
\[
    \alpha = \begin{dcases}
        O(w^{1/2}\eta^{1/p}\gamma^{-1/2})\bracks*{\parens*{\parens*{\log\frac{\tau n}{\eps}}^{1/2}\log d}^{1+1/l} + \sqrt l} & p < 2 \\
        O(\eps w^{1/2} \tau^{1/2} R)\bracks*{\parens*{\parens*{\log\frac{\tau n}{\eps}}^{1/2}\log d}^{1+1/l} + \sqrt l} & p > 2
    \end{dcases}
\]
We now set $\alpha = \eps$ and solve for the $\eps$ that we can obtain. From this, we see that we can set
\[
    \eps = \begin{dcases}
        O(w\eta^{2/p})^{1/2} \gamma^{-1/2}\bracks*{\parens*{(\log d)^2\log n}^{1+1/l} + l}^{1/2} & p < 2 \\
        O(w\eta \norm{\bfw}_1^{p/2-1})^{1/p}\gamma^{-1/2}\bracks*{\parens*{(\log d)^2\log n}^{1+1/l} + l}^{1/p} & p > 2
    \end{dcases}.
\]

\end{proof}

\section{Missing proofs for weak coresets}
\label{sec:weak-coreset-proofs}

\subsection{Proof of the closeness lemma}

\begin{proof}[Proof of Lemma \ref{lem:closeness}]
First note that
\begin{align*}
    \angle*{(\bfA\bfX^*\bfG - \bfB)^{\circ(p-1)}, \bfA\bfX^*\bfG-\bfA\bfX\bfG} &= \sum_{i=1}^n \sum_{j=1}^m [\bfA\bfX^*\bfG - \bfB](i,j)^{\circ(p-1)}[\bfA(\bfX^*-\bfX)\bfG](i,j) \\
    &= \sum_{i=1}^n \sum_{j=1}^m [\bfA\bfX^*\bfG - \bfB](i,j)^{\circ(p-1)}\angle*{(\bfA^\top\bfe_i)(\bfe_j^\top\bfG^\top),\bfX^*-\bfX} \\
    &= \angle*{\sum_{i=1}^n \sum_{j=1}^m [\bfA\bfX^*\bfG - \bfB](i,j)^{\circ(p-1)}(\bfA^\top\bfe_i)(\bfe_j^\top\bfG^\top),\bfX^*-\bfX}.
\end{align*}
The left term in the product is the gradient of the objective at the optimum by Lemma \ref{lem:grad-multiple-regression}, so this is just $0$ for any $\bfX$. Then for $p<2$, we have by Lemma \ref{lem:closeness-vec} that
\begin{align*}
    \norm{\bfA\bfX^*\bfG-\bfB}_{p,p}^2 + \frac{p-1}{2}\norm{\bfA\bfX\bfG-\bfA\bfX^*\bfG}_{p,p}^2 \leq \norm{\bfA\bfX\bfG-\bfB}_{p,p}^2 \leq (1+\eta)^2 \norm{\bfA\bfX^*\bfG-\bfB}_{p,p}^2
\end{align*}
which rearranges to
\[
    \norm{\bfA\bfX\bfG-\bfA\bfX^*\bfG}_{p,p} \leq O(\eta^{1/2})\OPT.
\]
and for $p > 2$, we have by Lemma \ref{lem:closeness-vec} that
\begin{align*}
    \norm{\bfA\bfX^*\bfG-\bfB}_{p,p}^p + \frac{p-1}{p2^p}\norm{\bfA\bfX\bfG-\bfA\bfX^*\bfG}_{p,p}^p &\leq \norm{\bfA\bfX\bfG-\bfB}_{p,p}^p \leq (1+\eta)^p \norm{\bfA\bfX^*\bfG-\bfB}_{p,p}^p
\end{align*}
which rearranges to
\[
    \norm{\bfA\bfX\bfG-\bfA\bfX^*\bfG}_{p,p} \leq O(\eta^{1/p})\OPT.
\]
\end{proof}

\subsection{Proof of the initial weak coreset bound}

\begin{proof}[Proof of Lemma \ref{lem:initial-iteration}]
We first show that
\[
    \norm{\bfA\hat\bfX\bfG - \bfA\bfX^*\bfG}_{p,p}^p \leq O\parens*{\frac1\delta} \OPT^p
\]
with probability at least $1-\delta$. By using the fact that $\bfS$ is an $O(1)$-approximate $\ell_p$ subspace embedding, we have that
\begin{align*}
    \norm{\bfA\hat\bfX\bfG - \bfA\bfX^*\bfG}_{p,p}^p &\leq \norm{\bfS(\bfA\hat\bfX\bfG - \bfA\bfX^*\bfG)}_{p,p}^p \\
    &\leq 2^{p-1}\parens*{\norm{\bfS(\bfA\hat\bfX\bfG-\bfB)}_{p,p}^p + \norm*{\bfS(\bfA\bfX^*\bfG-\bfB)}_{p,p}^p} && \text{Fact \ref{fact:tri}} \\
    &\leq 2^{p+1} \norm*{\bfS(\bfA\bfX^*\bfG-\bfB)}_{p,p}^p && \text{Approximate optimality of $\hat\bfX$}
\end{align*}
The latter quantity is at most $O(\frac1\delta)\OPT^p$ with probability at least $1-\delta$ by Markov's inequality. Thus, we may replace the optimization of $\hat\bfX$ over all $\bfX\in\mathbb R^{d\times t}$ with optimization over the ball $\braces{\bfX:\norm{\bfA\bfX\bfG - \bfA\bfX^*\bfG}_{p,p}^p = O(\frac1\delta)\OPT^p}$.

The rest of the proof now mimics the proof of Theorem \ref{thm:strong-coreset}. We apply Theorem \ref{thm:lewis-weight-sampling-diff-cor} with accuracy parameter $\eps$ set to $\eps\delta$, failure parameter set to $(\eps\delta)^p \delta^2$, and proximity parameter $\eta$ set to $1$. Let $S\subseteq[m]$ be the set of columns for which Theorem \ref{thm:lewis-weight-sampling-diff-cor} fails. Then by applying Markov's inequality twice as in the proof of Theorem \ref{thm:strong-coreset}, we have that
\[
    \sum_{j\in S}\norm{\bfS(\bfA\bfX^*\bfG-\bfB)\bfe_j}_p^p = O((\eps\delta)^p) \OPT^p
\]
and
\[
    \sum_{j\in S}\norm{(\bfA\bfX^*\bfG-\bfB)\bfe_j}_p^p = O((\eps\delta)^p) \OPT^p
\]
and thus it follows that
\[
    \sum_{j\in S}\norm{\bfS(\bfA\bfX\bfG-\bfB)\bfe_j}_p^p = \sum_{j\in S}\norm{(\bfA\bfX\bfG-\bfB)\bfe_j}_p^p \pm O(\eps\delta) \parens*{\norm{\bfA(\bfX-\bfX^*)\bfG}_p^p + \OPT^p}.
\]
Summing this result with the rest of the columns $j\notin S$ gives that
\begin{align*}
    &\abs*{\parens*{\norm{\bfS(\bfA\bfX\bfG-\bfB)}_{p,p}^p - \norm{\bfS(\bfA\bfX^*\bfG-\bfB)}_{p,p}^p} - \parens*{\norm{\bfA\bfX\bfG-\bfB}_{p,p}^p - \norm{\bfA\bfX^*\bfG-\bfB}_{p,p}^p}} \\
    \leq~& \eps\delta \parens*{\norm{\bfA\bfX^*\bfG-\bfB}_{p,p}^p + \norm{\bfS(\bfA\bfX^*\bfG-\bfB)}_{p,p}^p + \norm{\bfA\bfX\bfG-\bfA\bfX^*\bfG}_{p,p}^p} \leq O(\eps)\OPT^p
\end{align*}
Thus, in the ball $\braces{\bfX:\norm{\bfA\bfX\bfG - \bfA\bfX^*\bfG}_{p,p}^p = O(\frac1\delta)\OPT^p}$, we have that
\[
    \norm{\bfS(\bfA\bfX\bfG-\bfB)}_{p,p}^p = \norm{\bfA\bfX\bfG-\bfB}_{p,p}^p + (\norm{\bfS(\bfA\bfX^*\bfG-\bfB)}_{p,p}^p - \norm{\bfA\bfX^*\bfG-\bfB}_{p,p}^p) \pm O(\eps)\OPT^p.
\]
It follows that $\hat\bfX$ must minimize $\norm{\bfA\bfX\bfG-\bfB}_{p,p}^p$ up to an additive $O(\eps)\OPT^p$.
\end{proof}

\subsection{Proof of the weak coreset construction}

\begin{proof}[Proof of Theorem \ref{thm:weak-coreset-multiple-regression}]
Let
\[
    C = \begin{dcases}
        O(\gamma^{-1}) \delta^{-2}\norm{\bfw}_1 \bracks*{(\log d)^2\log n + \log\frac1\delta} & p < 2 \\
        O(\gamma^{-p/2}) \delta^{-p}\norm{\bfw}_1^{p/2}\bracks*{(\log d)^2\log n + \log\frac1\delta} & p > 2
    \end{dcases}
\]
We will make use of the fact that $\norm*{\bfS(\bfA\bfX^*\bfG-\bfB)}_{p,p}^p = O(\frac1\delta)\norm*{\bfS(\bfA\bfX^*\bfG-\bfB)}_{p,p}^p$ with probability at least $1-\delta$ by Markov's inequality.

We will first give the argument for $p<2$. Suppose that $C/\eps^\beta$ rows are needed for a $(1+\eps)$-approximate weak coreset. Now choose $a$ such that $a-2 = -a\beta$, that is, $a = 2/(1 + \beta)$. Then for $\eta^{2/p} = \eps^a$, $C\eta^{2/p}/(\eps\delta)^2 = C/\eta^{(2/p) \beta}$ rows yields a $(1+\eta^{2/p})$-approximate weak coreset. Then, a $(1+\eta^{2/p})$-approximate minimizer $\bfX$ satisfies
\[
    \norm{\bfA\bfX\bfG-\bfA\bfX^*\bfG}_{p,p}^p \leq O(\eta) \norm{\bfA\bfX^*\bfG-\bfB}_{p,p}^p
\]
by Lemma \ref{lem:closeness}. For all such $\bfX$, an argument as done in Theorem \ref{thm:strong-coreset} and Lemma \ref{lem:initial-iteration} shows that $\norm{\bfS(\bfA\bfX\bfG-\bfB)}_{p,p}^p - \norm{\bfS(\bfA\bfX^*\bfG-\bfB)}_{p,p}^p$ and $\norm{\bfA\bfX\bfG-\bfB}_{p,p}^p - \norm{\bfA\bfX^*\bfG-\bfB}_{p,p}^p$ are close up to an additive error of
\[
    \eps\delta\parens*{\norm*{\bfA\bfX^*\bfG-\bfB}_{p,p}^p + \norm*{\bfS(\bfA\bfX^*\bfG-\bfB)}_{p,p}^p + \frac1\eta \norm*{\bfA\bfX\bfG-\bfA\bfX^*\bfG}_{p,p}^p} = O(\eps)\norm{\bfA\bfX^*\bfG-\bfB}_{p,p}^p
\]
Thus, $C/\eta^{(2/p)\beta}$ rows in fact gives a $(1+O(\eps))$-approximate minimizer. That is, if $C/\eps^\beta$ rows is sufficient for $(1+\eps)$-approximation, then $C/\eta^{(2/p)\beta} = C/\eps^{a\beta} = C/\eps^{2\beta/(1+\beta)}$ rows is sufficient for $(1+\eps)$-approximation as well. We may now iterate this argument. Consider the sequence $\beta_i$ given by
\[
    \beta_0 = 2, \qquad \beta_{i+1} = \frac{2\beta_i}{1+\beta_i}.
\]
The solution to this recurrence is given by the following lemma, with $p = 2$:

\begin{Lemma}
\label{lem:active-recurrence}
Let $p>1$ and let $\{\beta_i\}_{i=0}^\infty$ be defined by the recurrence relation $\beta_0 = p$ and $\beta_{i+1} = p\beta_i / (1+\beta_i)$. Then,
\[
    \beta_i = \frac1{p^{-i}(p^{-1}-(p-1)^{-1}) + (p-1)^{-1}}
\]
\end{Lemma}
\begin{proof}
Note that $\frac1{\beta_{i+1}} = \frac1p \frac1{\beta_i} + \frac1p$ so the sequence $\{a_i\}_{i=0}^\infty$ given by $a_i = 1/\beta_i$ satisfies the linear recurrence $a_{i+1} = \frac1p a_i + \frac1p$. Note that this recurrence has the fixed point $a = 1/(p-1)$, so the sequence $a_i' = a_i - a$ satisfies $a_{i+1}' = \frac1p a_i'$, which gives, $a_i' = p^{-i}a_0'$. Thus, $a_i-a = p^{-i}(a_0-a)$ so
\begin{align*}
    \beta_i &= \frac1{a_i} = \frac1{p^{-i}(a_0-a) + a} \\
    &= \frac1{p^{-i}(p^{-1}-(p-1)^{-1}) + (p-1)^{-1}}.\qedhere
\end{align*}
\end{proof}

Thus, applying this argument $O(\log\log\frac1\eps)$ times yields that $\beta_i \leq 1 + O(1/\log(\frac1\eps))$ which means that reading only $O(1)C/\eps$ entries suffices. Union bounding over the success of the $O(\log\log\frac1\eps)$ rounds completes the argument.

Next, let $p>2$. Suppose that $C/\eps^\beta$ rows are needed for a $(1+\eps)$-approximate weak coreset. Now choose $a$ such that $a-p = -a\beta$, that is, $a = p/(1 + \beta)$. Then for $\eta = \eps^a$, $C\eta/\eps^p = C/\eta^{\beta}$ rows yields a $(1+\eta)$-approximate weak coreset. Then, a $(1+\eta)$-approximate minimizer $\bfX$ satisfies
\[
    \norm{\bfA\bfX\bfG-\bfA\bfX^*\bfG}_{p,p}^p \leq O(\eta) \norm{\bfA\bfX^*\bfG-\bfB}_{p,p}^p
\]
by Lemma \ref{lem:closeness}. For all such $\bfX$, an argument as done in Theorem \ref{thm:strong-coreset} and Lemma \ref{lem:initial-iteration} shows that $\norm{\bfS(\bfA\bfX\bfG-\bfB)}_{p,p}^p - \norm{\bfS(\bfA\bfX^*\bfG-\bfB)}_{p,p}^p$ and $\norm{\bfA\bfX\bfG-\bfB}_{p,p}^p - \norm{\bfA\bfX^*\bfG-\bfB}_{p,p}^p$ are close up to an additive error of
\[
    \eps\parens*{\norm*{\bfA\bfX^*\bfG-\bfB}_{p,p}^p + \frac1\eta \norm*{\bfA\bfX\bfG-\bfA\bfX^*\bfG}_{p,p}^p} = O(\eps)\norm{\bfA\bfX^*\bfG-\bfB}_{p,p}^p
\]
Thus, $C/\eta^{\beta}$ rows in fact gives a $(1+O(\eps))$-approximate minimizer. That is, if $C/\eps^\beta$ rows is sufficient for $(1+\eps)$-approximation, then $C/\eta^{\beta} = C/\eps^{a\beta} = C/\eps^{p\beta/(1+\beta)}$ rows is sufficient for $(1+\eps)$-approximation as well. We may now iterate this argument. Consider the sequence $\beta_i$ given by
\[
    \beta_1 = p, \qquad \beta_{i+1} = \frac{p\beta_i}{1+\beta_i}.
\]
Then by Lemma \ref{lem:active-recurrence}, applying this argument $O(\log\log\frac1\eps)$ times yields that $\beta_i \leq (p-1) + O(1/\log(\frac1\eps))$ which means that reading only $O(1)C/\eps^{p-1}$ entries suffices. Union bounding over the success of the $O(\log\log\frac1\eps)$ rounds completes the argument.
\end{proof}

\section{Missing proofs for applications}

\subsection{Sublinear algorithm for Euclidean power means}
\label{sec:power-means}

\PowerMeans*
\begin{proof}
We will assume without loss of generality that by reading $O(\log\frac1\delta)$ rows of $\bfB$, we can identify an $O(1)$-approximate solution $\hat\bfx$ (see, e.g., Section 3.1 of \cite{MMWY2022}). Thus by subtracting off this solution, we may assume that $\norm{\bfB}_{p,2}^p = O(\OPT^p)$.

We then use Dvoretzky's thoerem to embed this problem into the entrywise $\ell_p$ norm, so that
\[
    \norm{\mathbf{1}\bfx^\top - \bfB}_{p,2}^p = (1\pm\eps)\norm{\mathbf{1}\bfx^\top\bfG - \bfB\bfG}_{p,p}^p
\]
for every center $\bfx\in\mathbb R^d$. This is now in a form where we may apply our weak coreset results for multiple $\ell_p$ regression of Theorem \ref{thm:weak-coreset-informal}. Note that in this particular setting, the $\bfA$ matrix corresponds to the $n\times d$ all ones matrix with $d = 1$, and the $\ell_p$ Lewis weights can be taken to be uniform.

Now consider running $L = O(\log\frac1\delta)$ independent instances of the weak coreset algorithm, each which has the property that the algorithm makes at most
\begin{equation}
\label{eq:single-instance-query-bound}
    O(\eps^{-\rho})\parens*{\log\frac1\eps + \log\frac1\delta}
\end{equation}
queries for $\rho = 2$ for $p = 1$, $\rho = 1$ for $1 < p < 2$, and $\rho = p-1$ for $2 < p < \infty$, and that if $\norm{\bfS(\mathbf{1}(\bfx^*)^\top\bfG - \bfB\bfG)}_{p,p}^p = O(\norm{\mathbf{1}(\bfx^*)^\top\bfG - \bfB\bfG}_{p,p}^p)$ for the optimal solution $\bfx^*$, then it succeeds with probability at least $1-\delta/L$. By a union bound, this holds for all $L$ instances.

By Markov's inequality, each instance satisfies $\norm{\bfS\bfB\bfG}_{p,p}^p = O(\norm{\bfB\bfG}_{p,p}^p)$ with probability at least $9/10$, so at least $2/3$ of the $L$ instances must satisfy this bound with probability at least $1-\delta$. By Dvoretzky's theorem, this means that $\norm{\bfS\bfB}_{p,2}^p = O(\norm{\bfB}_{p,2}^p)$. Then, if we restrict our attention to the $(2/3)L$ instances with the smallest values of $\norm{\bfS\bfB}_{p,2}^p$, then all of these instances must output a correct $(1+\eps)$-approximately optimal solution, simultaneously with probability $1-\delta$. This gives a query bound of $L$ times \eqref{eq:single-instance-query-bound}.
\end{proof}

\subsection{Spanning coresets for \texorpdfstring{$\ell_p$}{lp} subspace approximation}
\label{sec:subspace-approx}

We show that weak coreset construction imply spanning sets for $\ell_p$ subspace approximation.

\SpanningCoreset*
\begin{proof}
By first computing a strong coreset of size $\poly(k/\eps)$ \cite{HV2020}, we can assume that $n, d = \poly(k/\eps)$. 

Let $\bfP = \bfV\bfV^\top$ be the rank $k$ projection that minimizes $\norm{\bfA\bfP-\bfA}_{p,2}^p$. Note then that
\[
    \min_{\bfX\in\mathbb R^{k\times d}}\norm{\bfA\bfV\bfX-\bfA}_{p,2}^p = \norm{\bfA\bfP-\bfA}_{p,2}^p.
\]
We then use Dvoretzky's theorem to embed this problem into the entrywise $\ell_p$ norm, so that
\[
    \norm{\bfA\bfV\bfX-\bfA}_{p,2}^p = (1\pm\eps)\norm{\bfA\bfV\bfX\bfG-\bfA\bfG}_{p,p}^p
\]
for every $\bfX\in\mathbb R^{k\times d}$, for some fixed $\bfG\in\mathbb R^{d\times m}$ with $m = \poly(d/\eps)$. Then by our weak coreset result for multiple $\ell_p$ regression (Theorem \ref{thm:weak-coreset-multiple-regression}), there is a diagonal matrix $\bfS$ with
\[
    \nnz(\bfS) \leq \begin{cases}
        O(\eps^{-2}k)(\log(k/\eps))^3 & p = 1 \\
        O(\eps^{-1}k)(\log(k/\eps))^3 & 1 < p \leq 2 \\
        O(\eps^{1-p}k^{p/2})(\log(k/\eps))^3 & 2 < p < \infty \\
    \end{cases}
\]
such that any $(1+\eps)$-approximate minimizer $\hat\bfX$ of $\norm{\bfS(\bfA\bfV\bfX\bfG-\bfA\bfG)}_{p,p}^p$ satisfies
\[
    \norm{\bfA\bfV\hat\bfX\bfG-\bfA\bfG}_{p,p}^p \leq (1+\eps)\min_{\bfX\in\mathbb R^{k\times d}}\norm{\bfA\bfV\bfX\bfG-\bfA\bfG}_{p,p}^p.
\]
We will take $\hat\bfX$ to be
\[
    \hat\bfX = \arg\min_{\bfX\in\mathbb R^{k\times d}}\norm{\bfS(\bfA\bfV\bfX-\bfA)}_{p,2}^p
\]
which is indeed a $(1+\eps)$-approximate minimizer of $\norm{\bfS(\bfA\bfV\bfX\bfG-\bfA\bfG)}_{p,p}^p$ by Dvoretzky's theorem. Then, again by Dvoretzky's theorem, we then have for this $\hat\bfX$ that
\begin{align*}
    \norm{\bfA\bfV\hat\bfX-\bfA}_{p,2}^p &\leq (1+O(\eps))\min_{\bfX\in\mathbb R^{k\times d}}\norm{\bfA\bfV\bfX-\bfA}_{p,2}^p \\
    &= (1+O(\eps))\norm{\bfA\bfP-\bfA}_{p,2}^p.
\end{align*}
Finally, note that $\hat\bfX$ has row span contained in the row span of $\bfS\bfA$, since otherwise $\norm{\bfS(\bfA\bfV\bfX-\bfA)}_{p,2}^p$ can be reduced by projecting the rows of $\bfX$ onto $\rowspan(\bfS\bfA)$. Then, if $\bfP_F$ is the projection matrix onto $F = \rowspan(\hat\bfX)$, then for each row $i\in[n]$ of $\bfA$,
\[
    \norm{\bfP_F\bfa_i - \bfa_i}_2 = \min_{\bfx\in F}\norm{\bfx - \bfa_i}_2 \leq \norm{\hat\bfX^\top \bfV^\top\bfa_i - \bfa_i}_2
\]
so
\[
    \norm{\bfA\bfP_F-\bfA}_{p,2}^p \leq \norm{\bfA\bfV\hat\bfX-\bfA}_{p,2}^p.
\]
We thus conclude that there is a rank $k$ subspace in the row span of $\bfS\bfA$ that is $(1+\eps)$-approximately optimal.
\end{proof}

\section{Missing proofs for coreset lower bounds}
\label{sec:lower-bounds-proofs}

We provide missing proofs from Section \ref{sec:lower-bounds}.

We will use the following lemma from coding theory.

\begin{Theorem}[\cite{PTB2013}]
\label{thm:coding-random-points}
For any $p\geq 1$ and $d = 2^k - 1$ for some integer $k$, there exists a set $S\subseteq\{-1,1\}^d$ and a constant $C_p$ depending only on $p$ which satisfy
\begin{itemize}
    \item $\abs*{S} = d^p$
    \item For any $s,t\in S$ such that $s\neq t$, $\abs*{\angle*{s,t}} \leq C_p\sqrt d$
\end{itemize}
\end{Theorem}

\subsection{Strong coresets}

\begin{proof}[Proof of Theorem \ref{thm:strong-coreset-lower-bound}]
Let $s = d^{p/2}$ and let $S\subseteq\{\pm1\}^d$ be a set of $\abs{S} = s$ points given by Theorem \ref{thm:coding-random-points} such that $\angle{\bfa,\bfa'} \leq C_{p/2} \sqrt d = O(\sqrt d)$ for some $C_{p/2}^p \geq 1$, for every distinct $\bfa, \bfa'\in S$. Let $m = s\eps^{-p}$, let $\bfA\in\{\pm1\}^{m\times d}$ be the matrix with $\eps^{-p}$ copies of $\bfa$ in its rows for each $\bfa\in S$, and let $\bfB = d\cdot \bfI_m$ be the $m\times m$ identity matrix scaled by $d$. For each row $i\in[m]$, we say that $i'\in[s]$ is its \emph{group number} if $\bfe_i^\top\bfA$ is the $i'$-th point in $S$.

Suppose for contradiction that $\bfS$ is a strong coreset with $\nnz(\bfS) \leq m/16$ such that
\[
    \norm{\bfS(\bfA\bfX-\bfB)}_{p,p}^p = \parens*{1\pm\frac{\eps}{12 C_{p/2}^p}}\norm{\bfA\bfX-\bfB}_{p,p}^p
\]
for every $\bfX\in\mathbb R^{d\times m}$. Then, there is a subset $T\subseteq[m]$ with $\abs{T} = m/16$ such that $\bfS$ is supported on $T$. For each $i'\in [s]$, let $T_{i'}\subseteq T$ denote the rows of $T$ whose rows in $\bfA$ with group number $i'\in[s]$, so $\sum_{i'=1}^s \abs{T_{i'}} = \abs{T}$. Then by averaging, there are at least $(3/4)s$ groups $i'\in[s]$ such that $\abs{T_{i'}} \leq \eps^{-p}/2$. Thus, we may assume without loss of generality that $\abs{T_{i'}} = \eps^{-p}$ for the first $(1/4)s$ groups, $\abs{T_{i'}} = \eps^{-p}/2$ for the last $(3/4)s$ groups, and $\abs{T} = (5/8)m$.

Let $W \coloneqq \sum_{i=1}^m \abs{\bfS_{i,i}}^p$ denote the total weight mass of $\bfS$. Note then that by querying $\bfX = 0$, we must have that
\[
    \norm{\bfS\bfB}_{p,p}^p = W = (1\pm\eps)\norm{\bfB}_{p,p}^p = \parens*{1\pm \frac{\eps}{12 C_{p/2}^p}} m.
\]
Let $W_1$ denote the sum of $\abs{\bfS_{i,i}}^p$ on the first $(1/4)s$ groups, and let $W_2$ denote the sum of $\abs{\bfS_{i,i}}^p$ on the last $(3/4)s$ groups. We will assume that $W_1 \leq m/4$, since the case of $W_1 \geq m/4$ is symmetric.

We now construct a query $\bfX\in\mathbb R^{d\times m}$ with the $j$-th column given by
\[
    \bfX\bfe_j = \begin{cases}
        \eps \cdot \bfe_j^\top\bfA & j\in T \\
        0 & j\notin T
    \end{cases}
\]
Note then that for each $i,j\in[m]$,
\[
    \bfe_i^\top\bfA\bfX\bfe_j = \begin{cases}
        \eps d & \bfe_i^\top\bfA = \bfe_j^\top \bfA, j\in T \\
        \eps C_{p/2}\sqrt d & \bfe_i^\top\bfA \neq \bfe_j^\top \bfA, j\in T \\
        0 & j\notin T
    \end{cases}
\]
Let $i\in[m]$ and let $i'\in[s]$ be its group number. Then the cost of row $i$ if $i\in T$ is
\begin{align*}
    \norm{\bfe_i^\top\bfA\bfX - \bfe_i^\top\bfB}_p^p = \sum_{j=1}^m \abs*{\bfe_i^\top\bfA\bfX\bfe_j - \bfB(i,j)}^p &= \underbrace{(1-\eps)^p d^p}_{i = j} + (\abs{T_{i'}}-1)\cdot \underbrace{\eps^p d^{p}}_{\bfe_i^\top\bfA = \bfe_j^\top\bfA} + (\abs{T} - \abs{T_{i'}})\cdot \underbrace{\eps^p C_{p/2}^p d^{p/2}}_{\bfe_i^\top\bfA \neq \bfe_j^\top\bfA} \\
    &= (1-p\eps + \abs{T_{i'}}\eps^p + (5/8)C_{p/2}^p + o(\eps))d^p
\end{align*}
while the cost of row $i\in[m]$ if $i\notin T$ is
\begin{align*}
    \norm{\bfe_i^\top\bfA\bfX - \bfe_i^\top\bfB}_p^p = \sum_{j=1}^m \abs*{\bfe_i^\top\bfA\bfX\bfe_j - \bfB(i,j)}^p &= \underbrace{d^p}_{i = j} + \abs{T_{i'}}\cdot \underbrace{\eps^p d^{p}}_{\bfe_i^\top\bfA = \bfe_j^\top\bfA} + (\abs{T} - \abs{T_{i'}})\cdot \underbrace{\eps^p C_{p/2}^p d^{p/2}}_{\bfe_i^\top\bfA \neq \bfe_j^\top\bfA} \\
    &= (1 + \abs{T_i'} \eps^p + (5/8)C_{p/2}^p + o(\eps))d^p.
\end{align*}
Let
\begin{align*}
    c_1 &= (1-p\eps + 1 + (5/8)C_{p/2}^p + o(\eps))d^p \\
    c_2 &= (1-p\eps + (1/2) + (5/8)C_{p/2}^p + o(\eps))d^p \\
    c_3 &= (1 + (1/2) + (5/8)C_{p/2}^p + o(\eps))d^p
\end{align*}
Then, the total true cost is at least
\begin{align*}
    \norm{\bfA\bfX-\bfB}_{p,p}^p &= \frac{m}{4}c_1 + \frac{3m}{8}c_2 +\frac{3m}{8}c_3 \\
    &= \frac{m}{4}c_1 + \frac{3m}{4}c_2 +\frac{3m}{8}(c_3 - c_2) \\
    &\geq \frac{m}{4}c_1 + \frac{3m}{4}c_2 + \frac{3m}{4}\cdot (\eps - o(\eps)) d^p
\end{align*}
while the strong coreset estimate is at most
\begin{align*}
    \norm{\bfS(\bfA\bfX-\bfB)}_{p,p}^p &= W_1 c_1 + W_2 c_2 \\
    &= W_1 (c_1-c_2) + (W_1 + W_2) c_2 \\
    &\leq \frac{m}{4} (c_1 - c_2) + \parens*{1+\frac{\eps}{12 C_{p/2}^p}} m c_2 \\
    &\leq \frac{m}{4} c_1 + \frac{3m}4 c_2 + \frac{\eps}{4} m d^p.
\end{align*}
Furthermore,
\[
    \frac{\eps}{12 C_{p/2}^p}\parens*{\frac{m}{4} c_1 + \frac{3m}4 c_2 + \frac{\eps}{4} m d^p} \leq \frac{\eps}{4} m d^p
\]
so $(1+\frac{\eps}{12 C_{p/2}^p})\norm{\bfS(\bfA\bfX-\bfB)}_{p,p}^p < \norm{\bfA\bfX-\bfB}_{p,p}^p$ and thus $\bfS$ fails to be a strong coreset. Rescaling $\eps$ by constant factors gives the desired result.
\end{proof}

\subsection{Weak coresets}

\begin{proof}[Proof of Theorem \ref{thm:weak-coreset-lower-bound}]
Our hard instance is identical to the one of Theorem \ref{thm:strong-coreset-lower-bound}, except that each group has $\eps^{1-p}/2C^p_{p/2}$ copies rather than $\eps^{-p}$ copies.

Note that if $\bfS$ does not sample some row $i\in[m]$, then the $i$-th column of $\bfS\bfB$ is all zeros, so the solution obtained by the weak coreset is $\bfX\bfe_i = 0$, which has objective function value $\norm{\bfB\bfe_i}_p^p = d^p$. On the other hand, the optimal value is at most $(1-\eps)^p d^p$ since we can set $\bfX\bfe_i = \eps \bfA^\top\bfe_i$ so that
\begin{align*}
    \norm{(\bfA\bfX-\bfB)\bfe_i}_p^p &\leq (1-\eps)^p d^p + \frac{\eps^{1-p}}{2C_{p/2}^p}\cdot \eps^p d^p + d^{p/2}\frac{\eps^{1-p}}{2C_{p/2}^p}\cdot C_{p/2}^p \eps^p d^{p/2} \\
    &\leq (1-\eps)^p d^p + \frac{\eps}{2}\cdot d^p + \frac{\eps}{2}\cdot d^{p} \\
    &\leq ((1-\eps)^p + \eps) d^p
\end{align*}
which is a $(1+\eps)$ factor smaller for all $\eps$ sufficiently small. Thus, if $\nnz(\bfS)\leq m/2$, then the solution $\bfX$ that minimizes $\norm{\bfS(\bfA\bfX-\bfB)}_{p,p}^p$ must be at least an additive $\eps d^p\cdot m/2$ more expensive than the optimal solution, and thus it fails to be a $(1+\eps/2)$-optimal solution.
\end{proof}

\subsection{Spanning coresets}

We generalize an argument of Section 4 of \cite{DV2006}.

\begin{Lemma}
\label{lem:rank-1-lower-bound}
Let $1 \leq p < \infty$ and
\[
    c_p = \begin{cases}
        1/6 & p \leq 2 \\
        1/(6\cdot 5^{p/2-1}) & p > 2
    \end{cases}
\]
Then, there is a matrix $\bfA\in\mathbb R^{n\times (n+1)}$ such that for every $\eps \geq 1/n$ and any subset of $s \leq c_p \eps^{-1}$ rows, any rank $1$ subspace $F'$ spanned by the $s$ rows must have
\[
    \norm{\bfA\bfP_{F'} - \bfA}_{p,2}^p > (1+\eps) \min_{\rank(F) \leq 1}\norm{\bfA\bfP_F - \bfA}_{p,2}^p.
\]
\end{Lemma}
\begin{proof}
Let $n \leq \eps^{-1}$ and let $\bfA$ be the $n\times (n+1)$ matrix given by $[R\cdot\mathbf{1}_n, \bfI_n]$ for some large enough $R>0$. That is, $\bfA$ is $R$ along the first column and the $n\times n$ identity for the last $n$ columns. Note that the optimal value is upper bounded by 
\[
    n((1-\eps)^2 + \eps^2\cdot (n-1))^{p/2} = n(1-2\eps + \eps^2 n)^{p/2} = n(1-\eps)^{p/2}.
\]

Let $\bfx\in\mathbb R^s$ be the coefficients of a linear combination of $s$ rows of $\bfA$. We may assume the coefficients are non-negative, since making the coefficients negative can only increase the cost. Note first that $1/2 \leq \norm{\bfx}_1 \leq 3/2$ since otherwise
\[
    n\cdot \abs*{R - R\norm{\bfx}_1}^p \geq n\cdot R/2
\]
which cannot be $(1+\eps)$-approximately optimal for $R\geq 2$.

The cost of the $i$-th row is $\parens*{(1-\bfx_i)^2 + \norm{\bfx}_2^2 - \bfx_i^2}^{p/2} = \parens*{1-2\bfx_i + \norm{\bfx}_2^2}^{p/2}$. If $\norm{\bfx}_2 \geq 2$, then
\[
    \parens*{1-2\bfx_i + \norm{\bfx}_2^2}^{p/2} \geq (1-2\norm{\bfx}_2 + \norm{\bfx}_2^2)^{p/2} = (\norm{\bfx}_2 - 1)^p \geq 1
\]
so this cannot produce a $(1+\eps)$-approximately optimal solution. Thus, assume $\norm{\bfx}_2 \leq 2$. Then,
\[
    \parens*{1-2\bfx_i + \norm{\bfx}_2^2}^{p/2} = \parens*{1 + \norm{\bfx}_2^2}^{p/2}\parens*{1-\frac{2}{1+\norm{\bfx}_2^2}\bfx_i}^{p/2} \geq \parens*{1 + \norm{\bfx}_2^2}^{p/2}\parens*{1-\frac{p}{1+\norm{\bfx}_2^2}\bfx_i}
\]
so summing over the rows gives a cost of
\begin{align*}
    \parens*{1 + \norm{\bfx}_2^2}^{p/2}\parens*{n-\frac{p}{1+\norm{\bfx}_2^2}\norm{\bfx}_1} &= \parens*{1 + \norm{\bfx}_2^2}^{p/2}n - p(1+\norm{\bfx}_2^2)^{p/2-1}\norm{\bfx}_1 \\
    &\geq \parens*{1 + \norm{\bfx}_1^2/s}^{p/2}n - p(1+\norm{\bfx}_2^2)^{p/2-1}\norm{\bfx}_1 && \text{since $1/2\leq \norm{\bfx}_1 \leq 3/2$} \\
    &\geq \parens*{1 + 1/2s}^{p/2}n - (3/2)p(1+\norm{\bfx}_2^2)^{p/2-1} \\
    &\geq \parens*{1 + p/4s}n - (3/2)p(1+\norm{\bfx}_2^2)^{p/2-1} \\
    &\geq \begin{cases}
        \parens*{1 + p/4s}n - (3/2)p & p \leq 2 \\
        \parens*{1 + p/4s}n - (3/2)p\cdot 5^{p/2-1} & p > 2 \\
    \end{cases}
\end{align*}
Thus, this fails to be a $(1+\eps)$-approximately optimal solution for
\[
    (p/4s)n \geq \begin{cases}
    (3/2)p & p \leq 2 \\
    (3/2)p\cdot 5^{p/2-1} & p > 2
    \end{cases}
\]
that is,
\[
    s \leq \begin{cases}
    n/6 & p \leq 2 \\
    n/(6\cdot 5^{p/2-1}) & p > 2
    \end{cases}.
\]
\end{proof}

We now extend Lemma \ref{lem:rank-1-lower-bound} to a general rank $k$ lower bound.

\begin{proof}[Proof of Theorem \ref{thm:rank-k-lower-bound}]
Let $n = \eps^{-1}$ and let $\bfB$ be a $kn\times k(n+1)$ block diagonal matrix with the $n\times(n+1)$ matrix construction $\bfA\in\mathbb R^{n\times (n+1)}$ of Lemma \ref{lem:rank-1-lower-bound} on the block diagonal. Consider any set $S$ of $s$ rows of $\bfB$, and let $S_i$ denote the set of $\abs{S_i} = s_i$ rows supported on the $i$-th block for each $i\in[k]$. Let $F_i$ denote the optimal subspace spanned by the rows $S_i$ on the $i$th block.

Let $T\subseteq [k]$ denote the set of $i\in[k]$ such that $s_i \leq c_p n$. If $i\in T$, then we by Lemma \ref{lem:rank-1-lower-bound} that
\[
    \norm{\bfA\bfP_{F_i} - \bfA}_{p,2}^p > \parens*{1+\frac{c_p}{s_i}} \min_{\rank(F) \leq k}\norm{\bfA\bfP_F - \bfA}_{p,2}^p
\]
Then, the additive error from these rows is bounded below by
\begin{align*}
    \sum_{i\in T}\frac{c_p}{s_i}\min_{\rank(F) \leq k}\norm{\bfA\bfP_F - \bfA}_{p,2}^p &\geq \abs{T}\cdot \frac{c_p \abs{T}}{\sum_{i\in[k] : s_i \leq c_p n} s_i}\min_{\rank(F) \leq k}\norm{\bfA\bfP_F - \bfA}_{p,2}^p && \text{AM-HM} \\
    &\geq \abs{T}\cdot \frac{c_p \abs{T}}{s}\min_{\rank(F) \leq k}\norm{\bfA\bfP_F - \bfA}_{p,2}^p \\
    &\geq \frac{c_p \abs{T}^2}{ks}\min_{\rank(F) \leq k}\norm{\bfB\bfP_F - \bfB}_{p,2}^p
\end{align*}
Note that $\abs{T} \geq k/2$ by averaging, so
\[
    \frac{c_p \abs{T}^2}{ks} \geq \frac{c_p k}{4s} \geq \eps
\]
which proves the theorem.
\end{proof}

\section{Experimental evaluation}

\begin{figure}[H]
\centering
\includegraphics[]{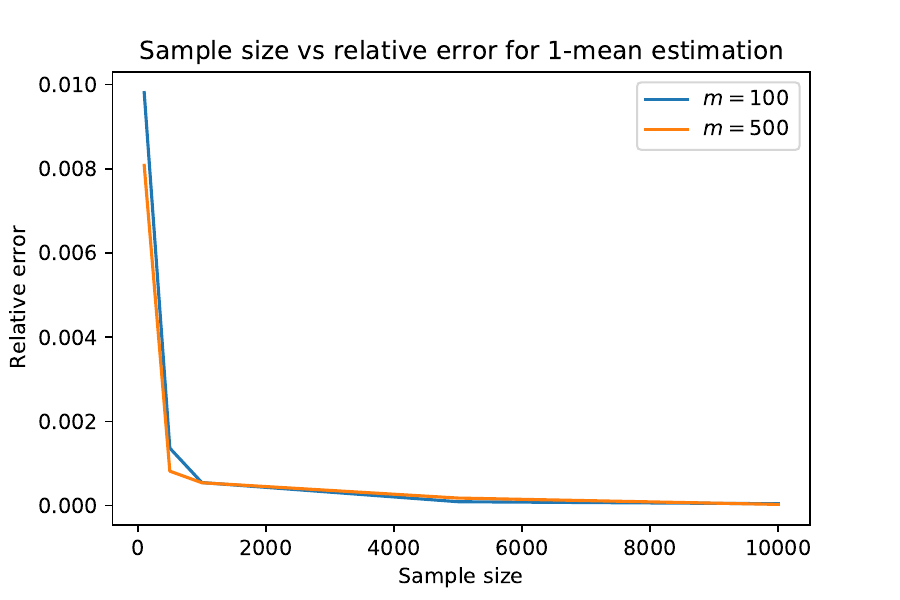}
\caption{Sample size vs relative error for $1$-mean estimation}
\label{fig:power-mean}
\end{figure}

We show that empirically, we indeed see that the trade-off between the number of uniform samples and the approximation quality is independent of the dimension $m$ in the setting of Euclidean power means. We do this by plotting the sample size against the resulting relative error for $m\in\{100,500\}$, where an $m$-dimensional dataset is constructed by sampling $m$ random features from the MNIST dataset. The results are shown in Figure \ref{fig:power-mean} and the experiment code is provided in Section \ref{sec:experiment-code}.

\subsection{Experiment code}
\label{sec:experiment-code}

We provide the code snippet for the experimental evaluation below.

\begin{python}
from keras.datasets import mnist
import numpy as np
import matplotlib.pyplot as plt
import tensorflow as tf

np.random.seed(2024)

(train_X, train_y), (test_X, test_y) = mnist.load_data()
train_X = train_X.reshape(len(train_X), -1)
train_X = train_X / np.max(train_X)
n, d = train_X.shape

def power_mean_loss(train_ds, x, p=1):
    x = np.expand_dims(x, axis=0)
    x = np.repeat(x, repeats=n, axis=0)
    e = train_ds - x
    e = np.linalg.norm(e, axis=-1)
    e = np.power(e, p)
    return np.sum(e) / n

def run(train_ds, max_iter=200, p=1):
    n, d = train_ds.shape
    x0 = np.zeros(d)
    x = tf.Variable(initial_value=x0)
    opt = tf.keras.optimizers.Adam(learning_rate=0.5)
    x.assign(x0)
    def power_mean_loss_tf():
        e = train_ds - x
        e = tf.norm(e, axis=-1)
        e = tf.math.pow(e, p)
        return tf.reduce_sum(e) / n
    losses = []
    while opt.iterations < max_iter:
        opt.minimize(power_mean_loss_tf, var_list=[x])
        loss = power_mean_loss_tf().numpy()
        if np.isnan(loss):
            print(x.numpy())
        losses.append(loss)
    return x.numpy(), losses

n, d = train_X.shape
sample_sizes = [100, 500, 1000, 5000, 10000]
for m in [100, 500]:
    cols = np.random.choice(d, m)
    train_m = train_X[:, cols]
    x, losses = run(train_m)
    OPT = losses[-1]
    estimates = []
    for sample_size in sample_sizes:
        train_sample = np.random.choice(n, sample_size)
        train_sample = train_m[train_sample, :]
        x, losses = run(train_sample)
        estimates.append(power_mean_loss(train_m, x))
    relative_errors = [(e / OPT) - 1 for e in estimates]
    print('relative errors', relative_errors)
\end{python}

\end{document}